\documentclass[twocolumn]{aastex631}

\def\p2{Pad$\rm \acute{e}_{(2,1)}$}
\def\LCDM{$\Lambda$CDM }

\usepackage{subfigure}

\usepackage{graphicx}
\usepackage{txfonts}

\usepackage{silence}
\begin{document}

\title{Hints of new physics for the Hubble tension: violation of cosmological principle}

\correspondingauthor{F. Y. Wang}
\email{fayinwang@nju.edu.cn}

\author[0000-0002-5819-5002]{J. P. Hu}
\affiliation{School of Astronomy and Space Science, Nanjing University, Nanjing 210093, China}

\author[0009-0009-3583-552X]{X. D. Jia}
\affiliation{School of Astronomy and Space Science, Nanjing University, Nanjing 210093, China}

\author[0000-0002-4797-4107]{J. Hu}
\affiliation{Institute of Astronomy and Information, Dali University, Dali 671003, China}

\author[0000-0003-4157-7714]{F. Y. Wang}
\affiliation{School of Astronomy and Space Science, Nanjing University, Nanjing 210093, China}
\affiliation{Key Laboratory of Modern Astronomy and Astrophysics (Nanjing University), Ministry of Education, Nanjing 210093, China}

\begin{abstract}
Discrepancy between the measurements of Hubble constant $H_{0}$ from the cosmic microwave background (CMB) and the local distance ladder is the most serious challenge to the standard $\Lambda$CDM model. Recent researches point out that it might be related with the violation of cosmological principle. Here, we investigate the impact of dipole-monopole correction on the constraints of $H_{0}$ utilizing the dipole fitting method based on the $\Lambda$CDM model and cosmography method. Our results show that the dipole-monopole correction can reduce the constraints of $H_{0}$ from a larger value consistent with SH0ES results to a smaller value consistent with Planck results. This finding can effectively alleviate the Hubble tension. Through making redshift tomography and model-independent analyses, we confirm that our findings are independent of redshift and cosmological model. In addition, the theoretical prediction of $H(z)/(1+z)$ reconstructed by the constraints of $\Lambda$CDM model with the dipole correction is in agreement with BAOs measurements including 5 DESI BAOs within 1$\sigma$ range except datapoint at z = 0.51. Our research suggests that the Hubble tension originates from new physics beyond the standard $\Lambda$CDM model, which might lead to a violation of the cosmological principle.
\end{abstract}


\section{Introduction} \label{sec:intro} 
Hubble tension is currently one of the most hot cosmological issues \citep{2020A&A...641A...6P,2022ApJ...934L...7R} which is the most serious challenge to the standard $\Lambda$CDM model. It stems from the huge discrepancy between the values of the Hubble constant ($H_{0}$) measured from the local distance ladder \citep[SH0ES;][]{2022ApJ...934L...7R} and the Planck cosmic microwave background \citep[CMB;][]{2020A&A...641A...6P}. The measurement of the former is 73.04$\pm$1.04 km/s/Mpc, and the ones of the latter is 67.36$\pm$0.54 km/s/Mpc. At present, the discrepancy has reached 5$\sigma$. To alleviate or resolve this tension, many schemes have been proposed, which can be roughly divided into three categories: (a) unknown systematic errors \citep{2019ApJ...876...85R,2022ApJ...934L...7R,2020A&A...641A...7P}; (b) independent observational arbitration \citep{2020MNRAS.498.1420W,2023A&A...673A...9S,2020ApJ...891L...1P,2018Natur.561..355M,2021ApJ...919...16F,2022MNRAS.515L...1W,2024MNRAS.527.7861G}; and (c) new physics beyond the standard cosmological model \citep{2021CQGra..38o3001D,2023Univ....9...94H}. Even so, there is currently no convincing solution. More information about the situation of the Hubble tension and its solutions can be obtained from the review articles \citep{2021CQGra..38o3001D,2022JHEAp..34...49A,2022NewAR..9501659P,2023Univ....9...94H,2023arXiv230911552K,2023Univ....9..393V,2024Univ...10..140C,2024IAUS..376...15R,2024MNRAS.527.7692W}. At present, research into the Hubble tension is still ongoing \citep{Scolnic:2024hbh,Toda:2024ncp}.

The Hubble tension is not the only challenge currently faced by the $\Lambda$CDM model, it also includes growth tension \citep{2017PhRvD..96f3517B,2018PhRvD..98d3526A,2018MNRAS.474.4894J} (2-3$\sigma$), Baryon Acoustic Oscillation (BAO) curiosities \citep{2017JCAP...04..024E,2018ApJ...853..119A,2019JCAP...10..044C} (2.5-3$\sigma$), small-scale curiosities \citep{2017ARA&A..55..343B,2019A&ARv..27....2S}, age of the universe \citep{2013PDU.....2..166V}, CMB anisotropy anomalies \citep{2020A&A...641A...7P} (2-3$\sigma$), violation of the cosmological principle \citep{2023CQGra..40i4001K} (2-5$\sigma$), $H_{0}$ descending trend \citep{2020PhRvD.102j3525K,2020MNRAS.498.1420W,2021ApJ...912..150D,2024arXiv240602019J} and so on. Furthermore, different types of observations seem to suggest that the Universe is anisotropic, including CMB \citep{2020A&A...641A...7P}, radio galaxies \citep{2018JCAP...04..031B,2018MNRAS.477.1772R}, galaxy cluster \citep{2020A&A...636A..15M,2021A&A...649A.151M,2024arXiv240601752M}, quasar \citep{2020A&A...643A..93H,2021ApJ...908L..51S,2021EPJC...81..694Z,2023ApJ...953..144G}, gamma-ray burst \citep[GRB;][]{2014MNRAS.443.1680W,2022MNRAS.511.5661Z}, type Ia supernovae \citep[SNe Ia;][]{2018MNRAS.474.3516W,2022A&A...668A..34H,2022PhRvD.105f3514K,2022PhRvD.105j3510L,2023PDU....3901162A,2023PhRvD.107b3507K,2023JCAP...07..020K,2023ChPhC..47l5101T,2024PhRvD.109l3533B}. Some recent researches pointed out that the violation of cosmological principle might be associated with the Hubble tension \citep{2023PDU....4201365Y,2024A&A...681A..88H,2024PDU....4601626Y}. For example, \citet{2023PDU....4201365Y} reported CMB independent constraints on $H_{0}$ in an anisotropic extension of $\Lambda$CDM model using a joint dataset including the Big Bang Nucleosynthesis (BBN), BAO, Cosmic Chronometer (CC), Pantheon+ sample and SH0ES Cepheid host distance anchors data. The analyses of the anisotropic model reveal that anisotropic level is positively correlated with $H_{0}$ values, and an anisotropy of the order $10^{-14}$ in the anisotropic model reduces the $H_{0}$ tension by $\sim$2$\sigma$. \citet{2024A&A...681A..88H} tested cosmological principle with the Pantheon+ sample and the region-fitting method and found relatively significant dipole anisotropy. Through redshift tomography analyses, they found that the local inhomogeneity might be closely related to the late-time $H_{0}$ transition \citep{2022MNRAS.517..576H} which can effectively alleviate the Hubble tension. 

In our present work, we will give $H_{0}$ constraints combining the Pantheon+ sample and the latest H(z) dataset employing different approaches. A redshift tomography analyses will be also made to test the redshift dependence of our findings. Afterwards, we will make a cross check utilizing a model independent method that is cosmography to analyse their model dependence. Finally, a brief comparison and discussion will be done combining the $H_{0}$ measurements from SH0ES and Planck collaborations, BAO and the Dark Energy Spectroscopic Instrument (DESI) BAO measurements. The outline of this paper is as follows. In Sect. \ref{sec2}, we give a basic information about the samples used, including redshift distribution and location distribution. Section \ref{sec3} introduces the $\Lambda$CDM model and the cosmography method in detail. The Markov chain Monte Carlo (MCMC) method and dipole fitting method are introduced in Sect. \ref{sec4}. In Sect. \ref{sec5}, we display the main results and give corresponding discussions. Finally, a brief summary is presented in Sect. \ref{sec6}.

\section{Dataset information} \label{sec2}
In this research, two types of observation were employed, including SNe Ia and observational Hubble parameter H(z). The SNe Ia comes from the Pantheon+ sample which is the largest sample at present. It consists of 1701 SNe Ia light curves observed from 1550 distinct sources and covers redshift range from 0.001 to 2.26 \citep{2022ApJ...938..110B,2022ApJ...938..113S}. This sample has been widely used for cosmological applications \citep{2023PhRvD.107j3521C,2023A&A...674A..45J,2023PDU....4201363Y,2024JCAP...08..015A,2024PhRvD.109l3533B,2024arXiv240600273C,2024PhRvD.110b3506P,2024ApJ...967...47L,2024PDU....4601626Y}. The H(z) dataset consists of 35 CCs including two parts: 31 old CCs \citep{2018ApJ...856....3Y} and 4 new CCs \citep{2022ApJ...928L...4B,2023ApJS..265...48J,2023JCAP...11..047J,2023AA...679A..96T}. The old dataset has been widely used in cosmological researches, such as cosmological constraints \citep{2018ApJ...856....3Y,2020CQGra..38e5007B,2020EPJC...80..562V}, calibration of the GRB correlations \citep{2015NewAR..67....1W,2021MNRAS.507..730H,2022ApJ...924...97W}, constraints of the transition redshift \citep{2016JCAP...05..014M,2020JCAP...04..053J}, cosmographic constraints \citep{2020MNRAS.491.4960L,2021PhRvD.103f3537A,2022BrJPh..52..115V} and alleviation of the Hubble tension \citep{2022MNRAS.517..576H}. 

\begin{figure}[htp]
	\centering
	\includegraphics[width=0.35\textwidth]{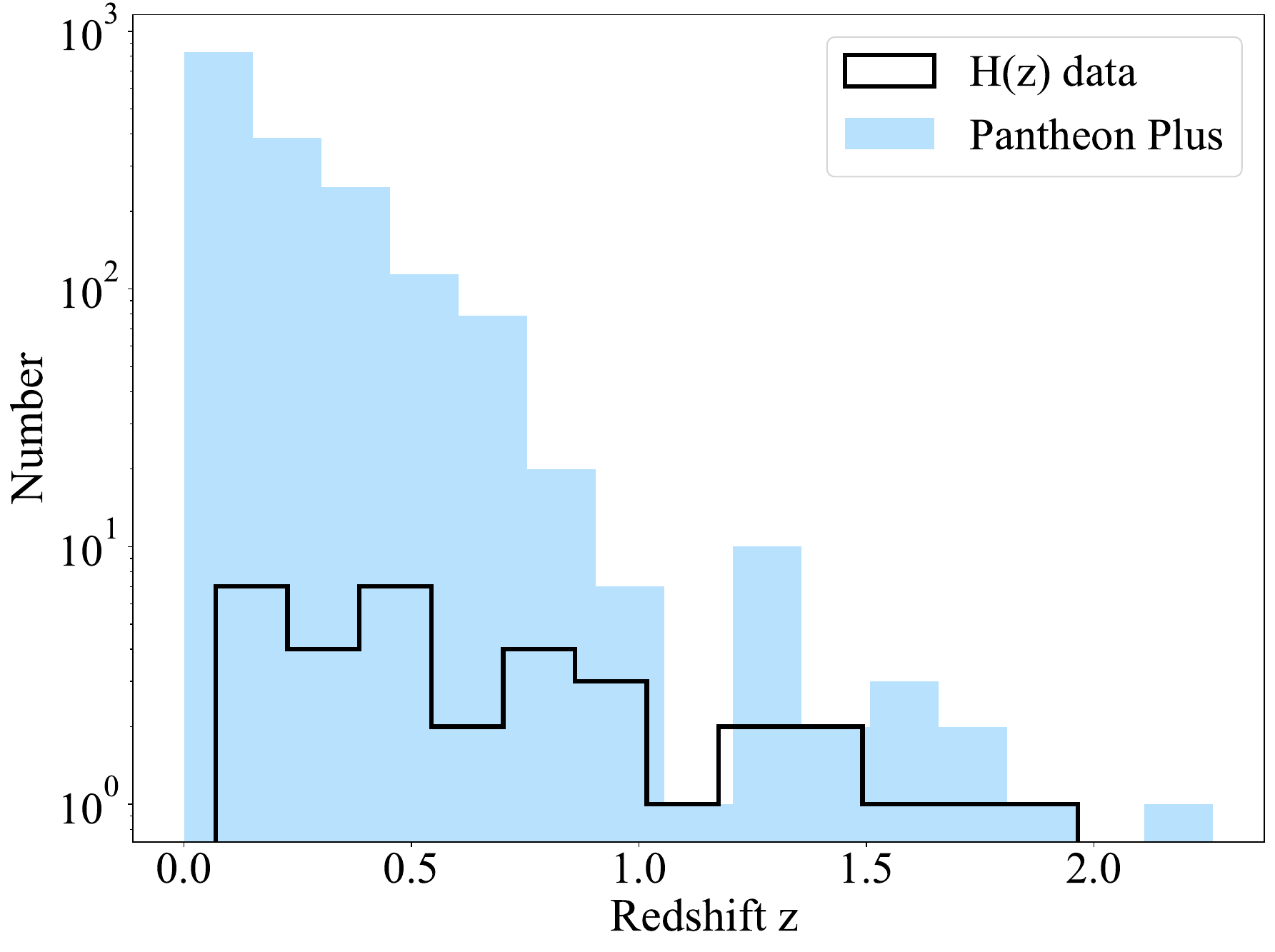}
	\caption{Redshift distribution of the Pantheon+ sample and the latest H(z) observation.}
	\label{rd}       
\end{figure}

\begin{figure}[htp]
	\centering
        \subfigure[0$<$ z $\leq$ 0.10]{\label{Fig9.sub.1}\includegraphics[width=0.49\hsize]{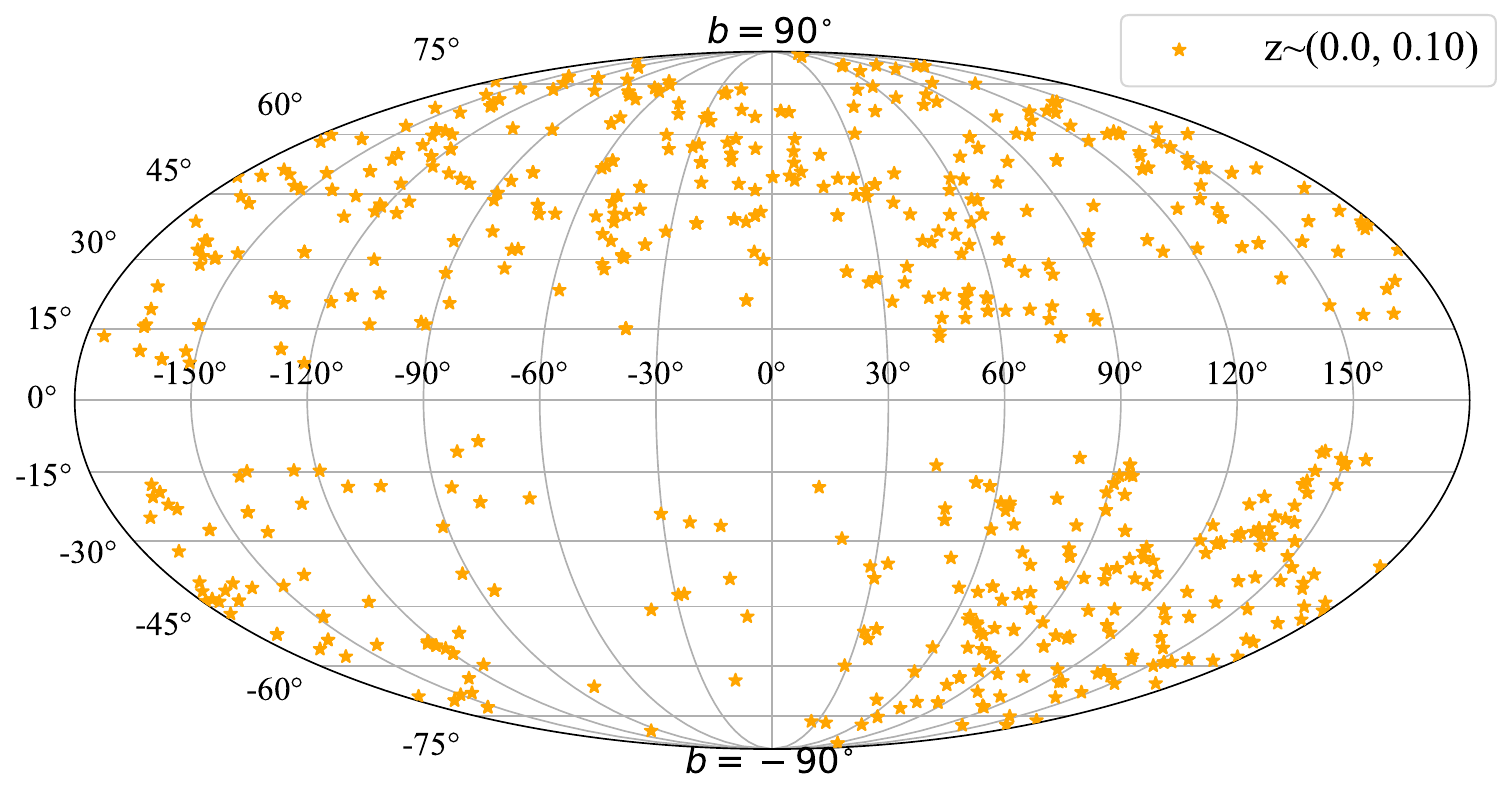}}
        \subfigure[0.10 $<$ z $\leq$ 0.20]{\label{Fig9.sub.2}\includegraphics[width=0.49\hsize]{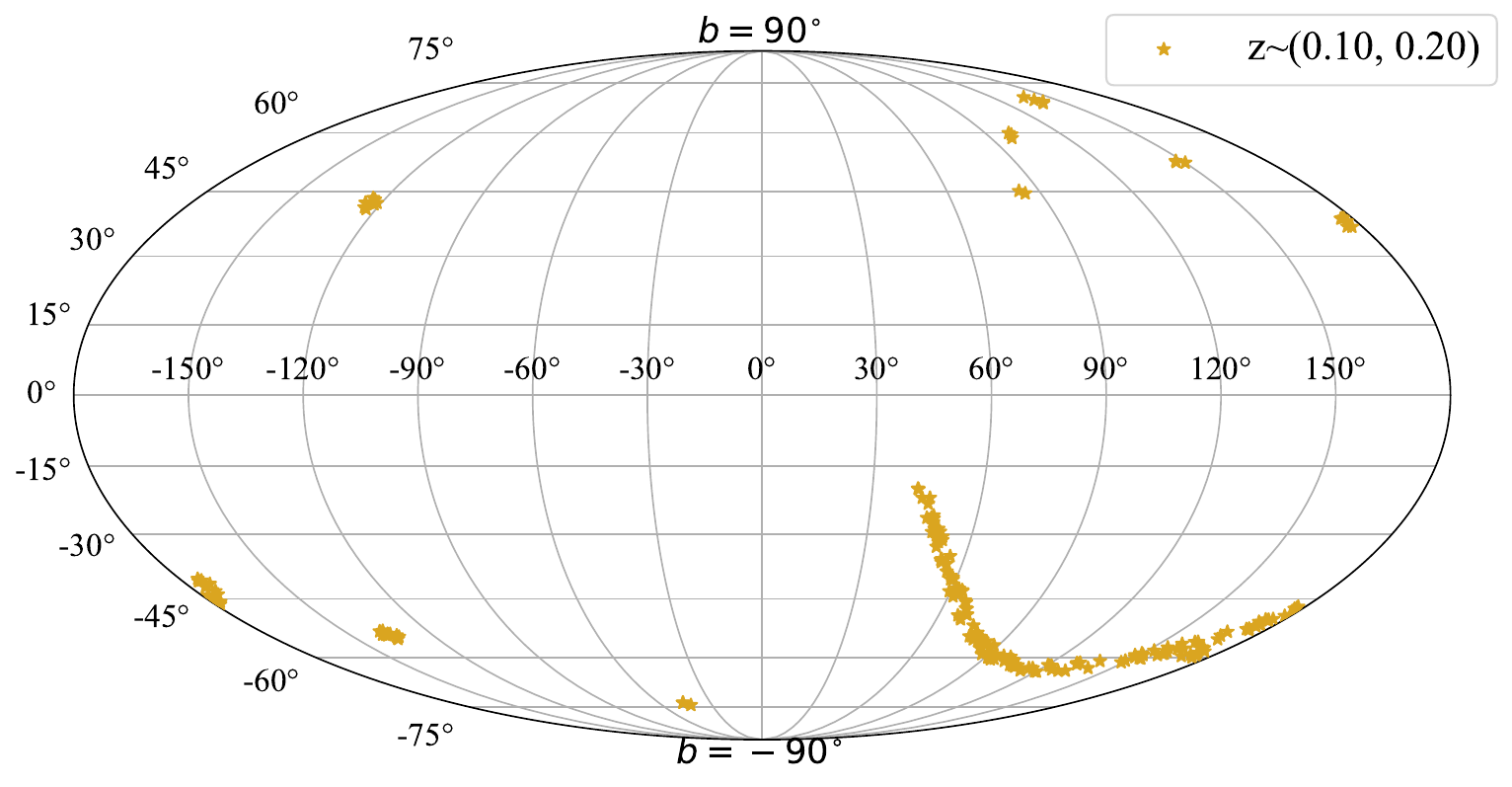}}\\
        \subfigure[0.20 $<$ z $\leq$ 0.30]{\label{Fig9.sub.3}\includegraphics[width=0.49\hsize]{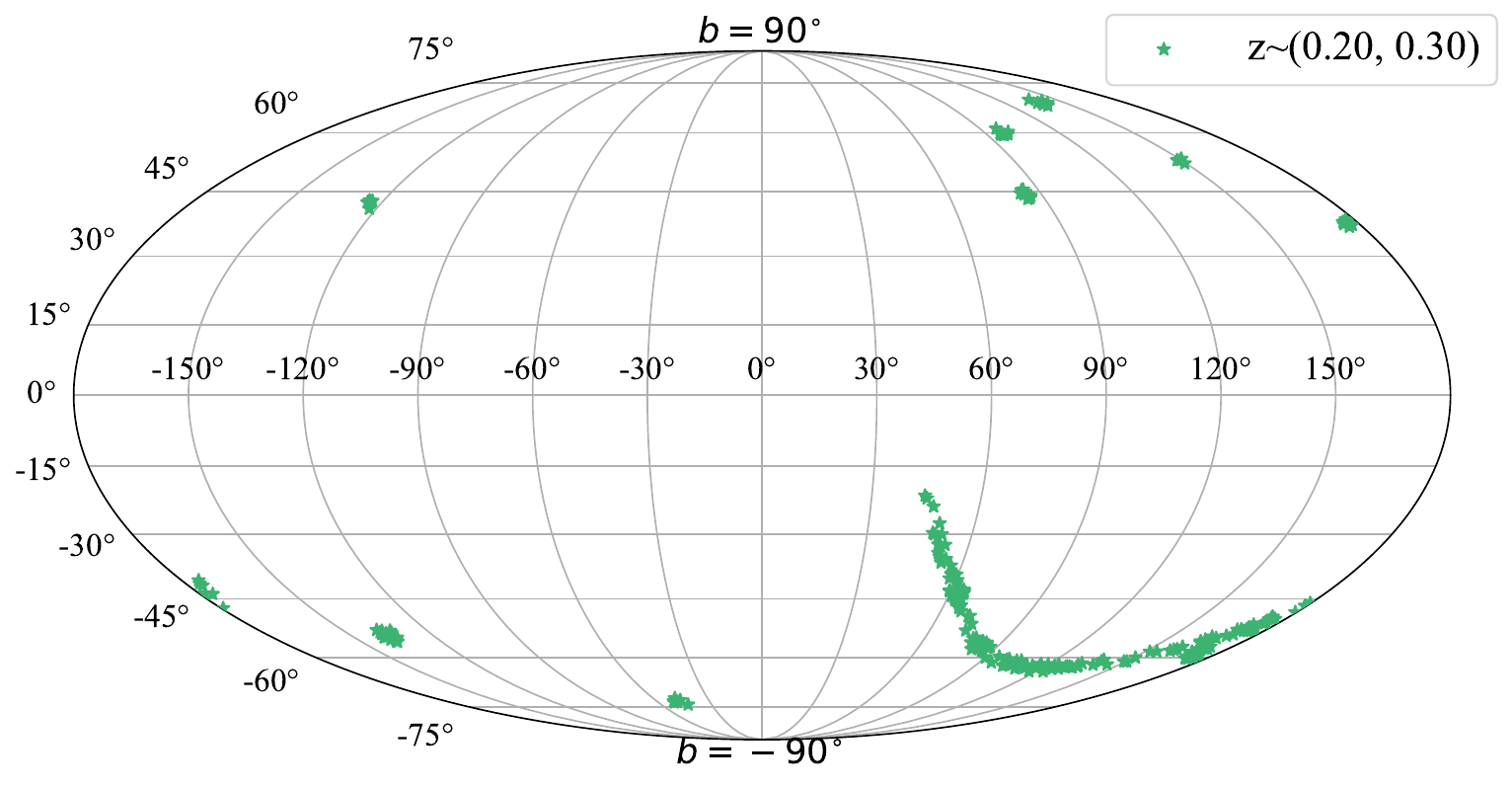}}
        \subfigure[0.30 $<$ z $\leq$ 0.50]{\label{Fig9.sub.4}\includegraphics[width=0.49\hsize]{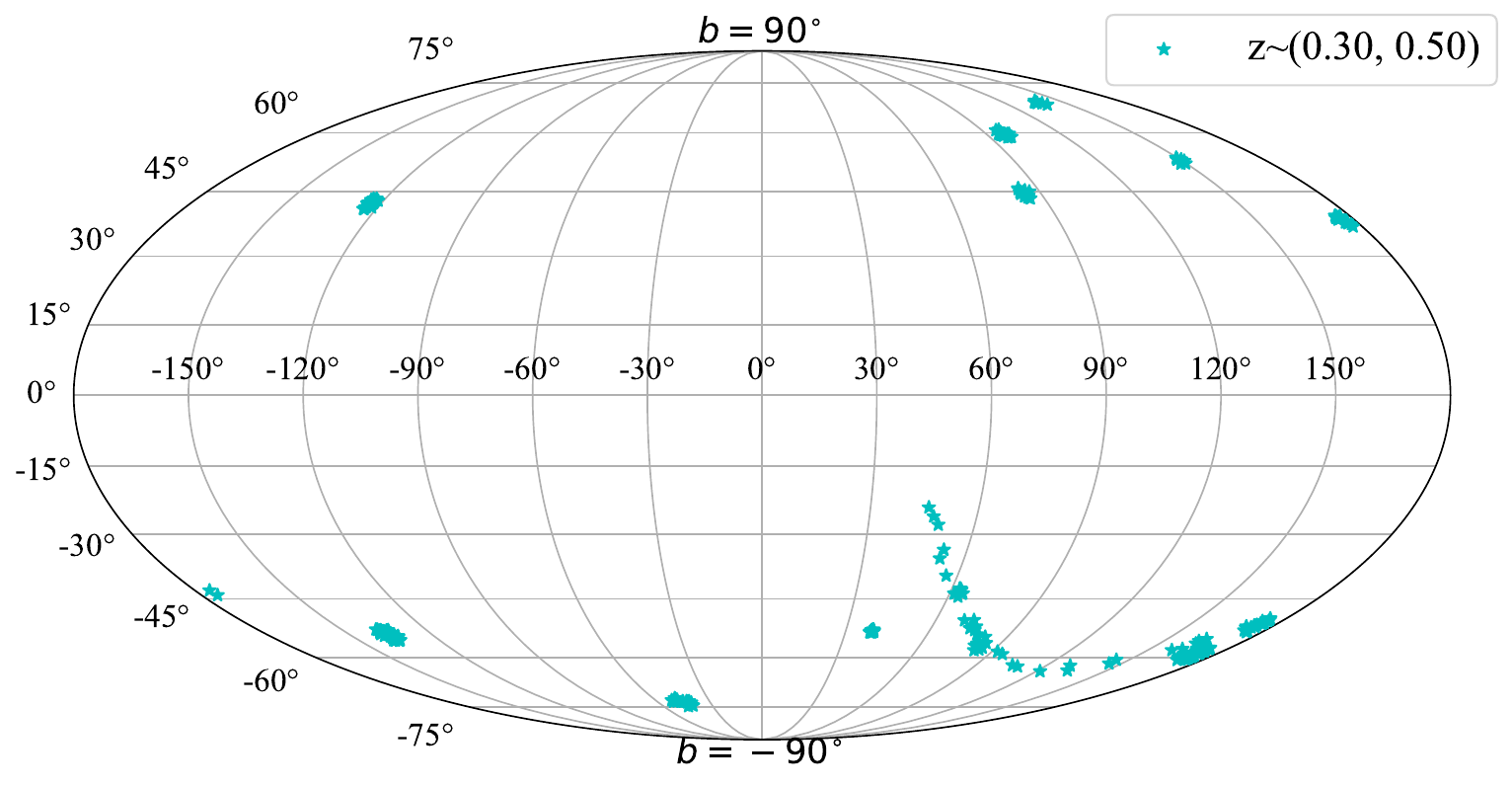}}\\
        \subfigure[0.50 $<$ z $\leq$ 2.26]{\label{Fig9.sub.5}\includegraphics[width=0.49\hsize]{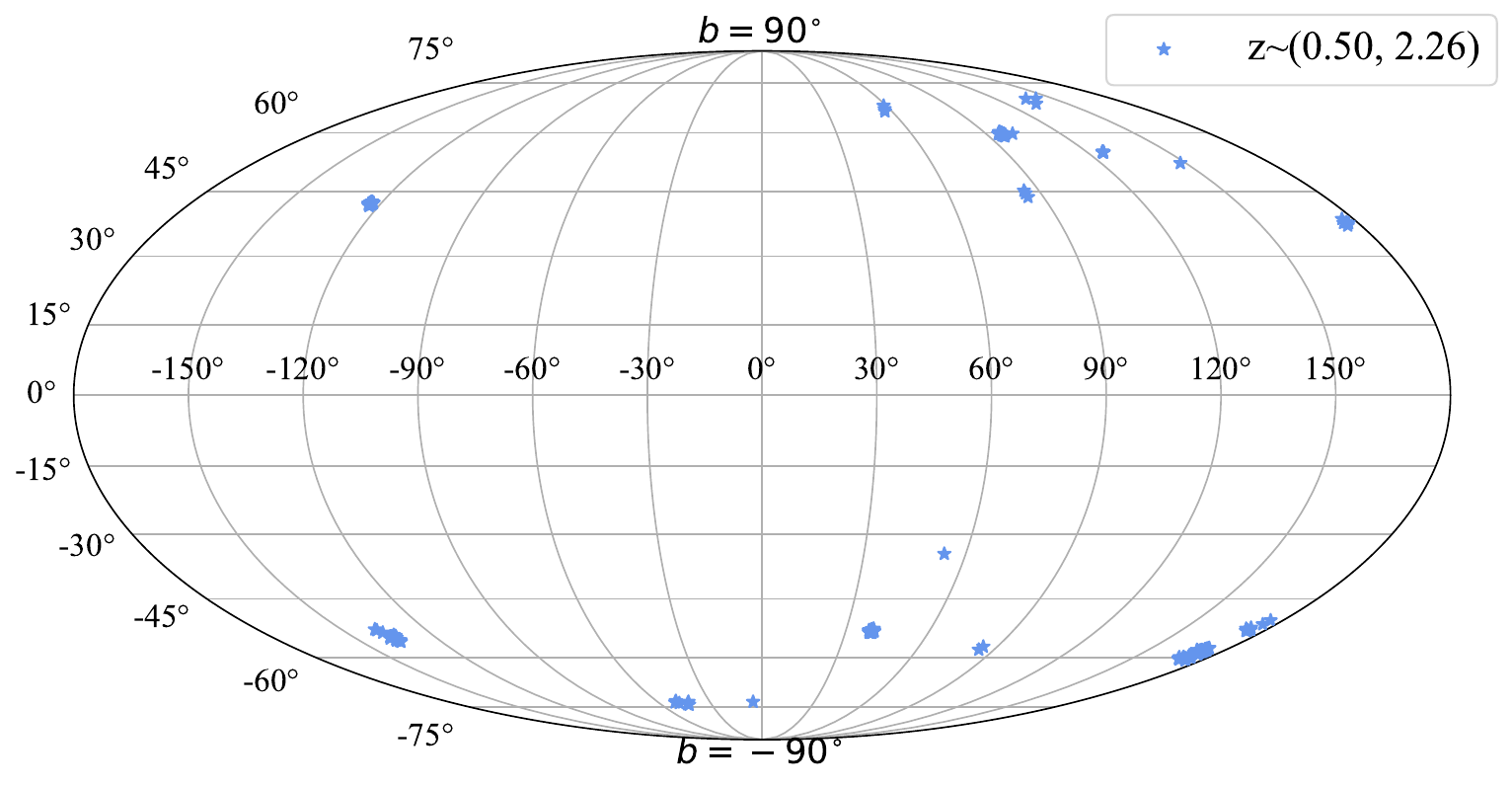}}
        \subfigure[Total sample]{\label{Fig9.sub.6}\includegraphics[width=0.49\hsize]{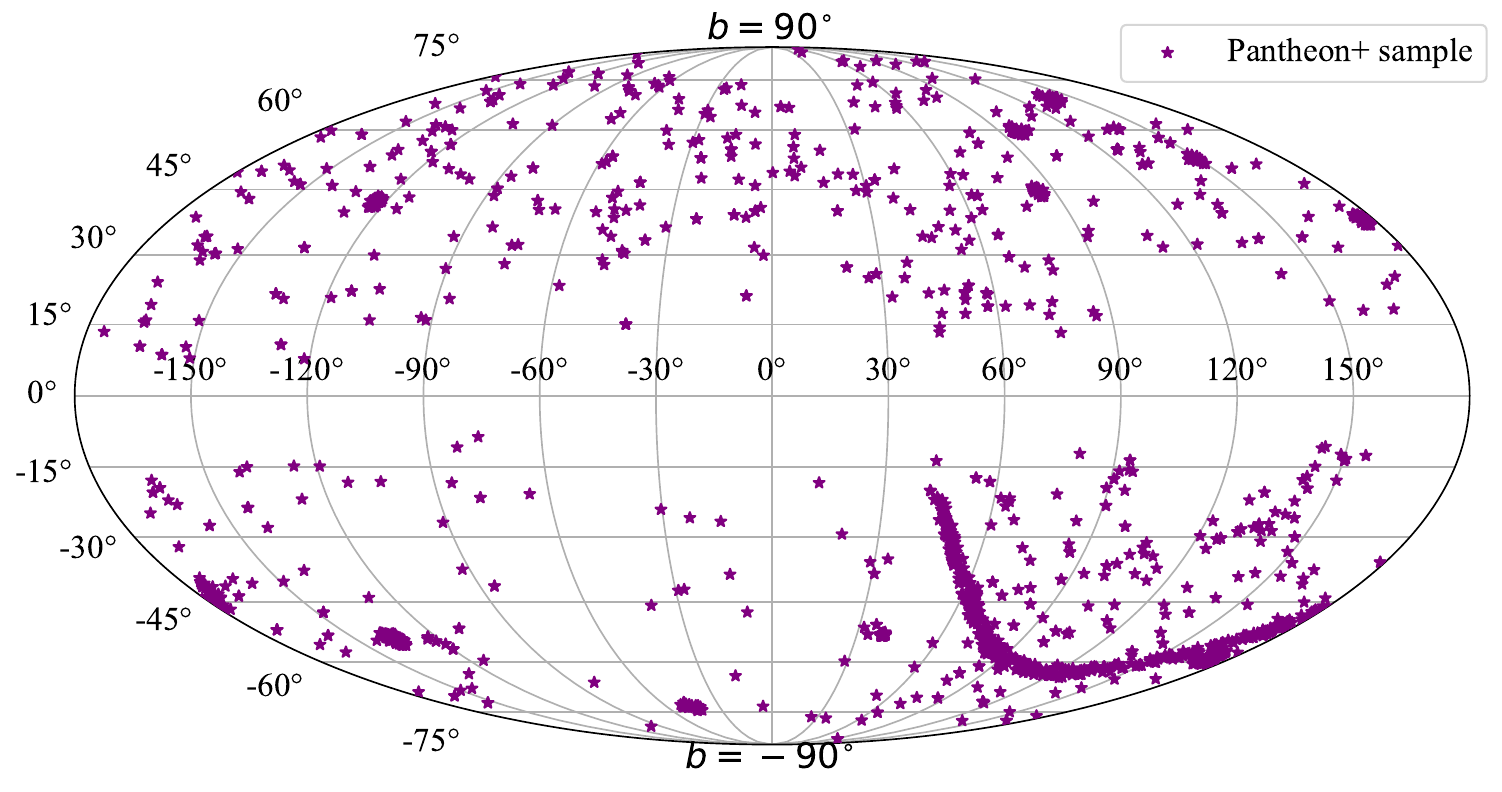}}
	\caption{Location distributions of different cut-off redshift SNe Ia in the galactic coordinate system. Panels (a)-(e) correspond to the redshift intervals (0, 0.10), (0.10, 0.20), (0.20, 0.30), (0.30, 0.50) and (0.50, 2.26), respectively. Panel (f) gives the location distribution of the Pantheon+ sample.} 
	\label{ld}       
\end{figure}

In Figure \ref{rd}, we gave the corresponding redshift distributions. From the redshift distribution, it is easy to find that nearly half of the SNe less than 0.10. We also mapped the location distribution of different redshift SNe in the galactic coordinate system, as shown in Figure \ref{ld}. It can be found that the spatial distribution of z $\leq$ 0.10 is relatively uniform, while the higher redshift data is relatively concentrated, mainly concentrated in a belt-like structure which observed by SDSS survey \citep{2018PASP..130f4002S}. For the H(z) dataset, it covers redshift range (0.07, 1.965). Detail information is shown in Table \ref{tab:CC}. From Figure \ref{rd}, we find its redshift distribution is relatively uniform below redshift 1.00, and there are relatively few data above 1.00, especially above 1.50.

\begin{table*}[htp]
	\caption{ $H(z)$ data from Cosmic Chronometer (in units of $\textrm{km}~\textrm{s}^{-1} \textrm{Mpc}^{-1}$). \label{tab:CC}}
	\centering
	\begin{tabular}{cccc|cccc}
		\hline\hline
   No.	&	$z$  & $H(z)$  & Reference & No.	&	$z$  & $H(z)$  & Reference    \\  
	\hline
  (1) 	&	$0.07$    & $69.0\pm19.6$ & \cite{2014RAA....14.1221Z}&(19)	&	$0.593$   & $104.0\pm13.0$ & \cite{2012JCAP...08..006M}\\
  (2) 	&	$0.09$    & $69.0\pm12.0$ & \cite{2005PhRvD..71l3001S}&(20)	&	$0.68$    & $92.0\pm8.0$  &  \cite{2012JCAP...08..006M}\\
  (3) 	&	$0.12$    & $68.6\pm26.2$ & \cite{2014RAA....14.1221Z}&(21)	&	$0.75$    & $98.8\pm33.6$  &  \cite{2022ApJ...928L...4B}\\
  (4)   &	$0.17$    & $83.0\pm8.0$ &  \cite{2005PhRvD..71l3001S}&(22)	&	$0.75$    & $105.0\pm7.9$  &  \cite{2023JCAP...11..047J}\\
  (5) 	&	$0.179$   & $75.0\pm4.0$ &  \cite{2012JCAP...08..006M}&(23)	&	$0.781$   & $105.0\pm12.0$ & \cite{2012JCAP...08..006M}\\
  (6) 	&	$0.199$   & $75.0\pm5.0$ &  \cite{2012JCAP...08..006M}&(24)	&	$0.80$    & $113.1\pm15.1$  &  \cite{2023ApJS..265...48J}\\
  (7) 	&	$0.2$     & $72.9\pm29.6$ & \cite{2014RAA....14.1221Z}& (25)	&	$0.875$   & $125.0\pm17.0$ & \cite{2012JCAP...08..006M}\\
  (8)	&	$0.27$    & $77.0\pm14.0$ & \cite{2005PhRvD..71l3001S}&(26)	&	$0.88$    & $90.0\pm40.0$  & \cite{2017MNRAS.467.3239R}\\
  (9)	&	$0.28$    & $88.8\pm36.6$ & \cite{2014RAA....14.1221Z}& (27)	&	$0.9$     & $117.0\pm23.0$ & \cite{2005PhRvD..71l3001S}\\
  (10)	&	$0.352$   & $83.0\pm14.0$ & \cite{2012JCAP...08..006M}& (28)	&	$1.037$   & $154.0\pm20.0$ & \cite{2012JCAP...08..006M}\\
  (11)	&	$0.3802$  & $83.0\pm13.5$ & \cite{2016JCAP...05..014M}& (29)	&	$1.26$    & $135.0\pm65$  &  \cite{2023AA...679A..96T}\\
  (12)	&	$0.4$     & $95.0\pm17.0$ & \cite{2005PhRvD..71l3001S}& (30)	&	$1.3$     & $168.0\pm17.0$ & \cite{2005PhRvD..71l3001S}\\
  (13)	&	$0.4004$  & $77.0\pm10.2$ & \cite{2016JCAP...05..014M}& (31)	&	$1.363$   & $160.0\pm33.6$ & \cite{2015MNRAS.450L..16M}\\
  (14)	&	$0.4247$  & $87.1\pm11.2$ & \cite{2016JCAP...05..014M}& (32)	&	$1.43$    & $177.0\pm18.0$ & \cite{2005PhRvD..71l3001S}\\
  (15)	&	$0.4497$ & $92.8\pm12.9$ &  \cite{2016JCAP...05..014M}& (33)	&	$1.53$    & $140.0\pm14.0$ & \cite{2005PhRvD..71l3001S}\\
  (16)	&	$0.47$    & $89.0\pm50$  &  \cite{2017MNRAS.467.3239R}& (34)	&	$1.75$    & $202.0\pm40.0$ & \cite{2005PhRvD..71l3001S}\\
  (17)	&	$0.4783$  & $80.9\pm9.0$ &  \cite{2016JCAP...05..014M}& (35)	&	$1.965$   & $186.5\pm50.4$ & \cite{2015MNRAS.450L..16M}\\
  (18)	&	$0.48$    & $97.0\pm62.0$ &  \cite{2017MNRAS.467.3239R}& & & & \\
		\hline\hline
	\end{tabular}
\end{table*}

\section{Models} \label{sec3}
Cosmological constraints require the theoretical forms of the distance modulus $\mu_{th}$ and the Hubble parameter $H(z)_{th}$ which need to assume a cosmological model. In the following content, we will give the corresponding theoretical expressions in the $\Lambda$CDM model and cosmography method.
\subsection{\rm{\LCDM} model}
The \LCDM model is the most simplest cosmological model with just two main ingredients which including the positive cosmological constant mimicking dark energy (DE) and the dark matter (DM). Considering a spatially flat $\Lambda$CDM model, the luminosity distance $d_{L}$ can be calculated from
\begin{equation}
d_{L} = \frac{c(1+z)}{H_{0}} \int_{0}^{z} 
\frac{dz'}{\sqrt{\Omega_{m} (1+z')^{3} + \Omega_{\Lambda}}},
\label{dl}
\end{equation}
where $c$ is the speed of light, $H_{0}$ is the Hubble constant,
$\Omega_{m}$ is the DM parameter, and $\Omega_{\Lambda}$ is the DE parameter, satisfying $\Omega_{m}$ + $\Omega_{\Lambda}$ = 1. Based on Equation (\ref{dl}), the theoretical distance modulus can be written as 
\begin{equation}
\mu_{th} = 5 \log_{10} \frac{d_{L}}{\textnormal{Mpc}} + 25.
\label{mu}
\end{equation}
The form of theoretical Hubble parameter ($H(z)_{th}$) under the spatially flat $\Lambda$CDM model can be written as 
\begin{equation}
H(z)_{th} = H_{0}\sqrt{\Omega_{m} (1+z)^{3} + \Omega_{\Lambda}}.
\label{hz}
\end{equation}

\subsection{Cosmography method}
Cosmography is a model-independent strategy which only relies on the assumption of a homogeneous and isotropic universe which is described by the Friedman-Lemaitre-Robertson-Walker (FLRW) metric well \citep{1972gcpa.book.....W}. It has been widely used to restrict the state of the kinematics of our Universe employing measured distances \citep{2015CQGra..32m5007V,2016IJGMM..1330002D,2019IJMPD..2830016C}. Its methodology is essentially based on expanding a measurable cosmological quantity into the Taylor series around the present time. This feature makes it can estimate the cosmic evolution at $z$ $\backsim$ 0 well, but failed at high redshifts \citep{1998tx19.confE.276C,2004JCAP...09..009C,2004ApJ...607..665R,2004CQGra..21.2603V,2022A&A...661A..71H}. In this framework, the evolution of the Universe can be described by some cosmographic parameters, such as Hubble parameter $H$, deceleration $q$, jerk $j$, snap $s$, and lerk $l$ parameters. The definition of them can be expressed as follows:
\begin{eqnarray}
        H = \frac{\dot{a}}{a}, q = -\frac{1}{H^{2}}\frac{\ddot{a}}{a}, j=\frac{1}{H^3}\frac{\dot{\ddot{a}}}{a}, s=\frac{1}{H^4}\frac{\ddot{\ddot{a}}}{a}, l=\frac{1}{H^5}\frac{\dot{\ddot{\ddot{a}}}}{a}.
\end{eqnarray} 
The Taylor Series expansion of the Hubble parameter around present time (z = 0) is
\begin{eqnarray}
        \label{eq:hz}
        H(z)_{th,T} &=& H_{0}[1 + (1+q_{0})z + \frac{1}{2}(j_{0} - q_{0}^{2})z^{2} - \frac{1}{6}(-3q_{0}^{2}- 3q_{0}^{3} \nonumber \\
        &+&j_{0}(3+4q_{0}) + s_{0})z^{3} +\frac{1}{24}(-4j_{0}^{2} + l_{0} - 12q_{0}^{2}-24q_{0}^{3} \nonumber \\
        &-&15q_{0}^{4}+j_{0}(12 + 32q_{0}+ 25q_{0}^{2})+ 8s_{0} + 7q_{0}s_{0})z^{4} \nonumber \\
        &+& \textit{O}(z^5)],
\end{eqnarray}
where $H_{0}$, $q_{0}$, $j_{0}$, $s_{0}$, and $l_{0}$ are the current values. The corresponding luminosity distance can be conveniently expressed as \citep{2007CQGra..24.5985C,2011PhRvD..84l4061C,2020MNRAS.494.2576C}
\begin{eqnarray}
        \label{eq:dlz}
        d_{L}(z)_{T} &=& \frac{c}{H_{0}}[z + \frac{1}{2} (1-q_{0})z^{2} -\frac{1}{6}(1-q_{0}-3q_{0}^{2}+j_{0})z^{3} \nonumber \\
        &+& \frac{1}{24}(2-2q_{0}-15q_{0}^{2}-15q_{0}^{3}+5j_{0}+10q_{0}j_{0}+s_{0})z^{4}\nonumber \\
        &+& \frac{1}{120}(-6+6q_{0}+81q_{0}^{2}+165q_{0}^{3}+105q_{0}^{4}+10j_{0}^{2} \nonumber \\
        &-&27j_{0}-110q_{0}j_{0}-105q_{0}^{2}j_{0}-15q_{0}s_{0}-11s_{0}-l_{0})z^{5} \nonumber \\ 
        &+&\textit{O}(z^6)].
\end{eqnarray} 
The first two terms of $d_{L}(z)_{T}$ are Weinberg's version of the Hubble law which can be found from Equation (14.6.8) in the book by \citet{1972gcpa.book.....W}. The third term and the fourth term are equivalent to that obtained by \citet{1998tx19.confE.276C} and \citet{2004CQGra..21.2603V}, respectively. In cosmography the radius of convergence of any z-expansion is z = 1.0. If the redshift range exceeds 1.0, there is no guarantee that one can approximate any model, including $\Lambda$CDM model \citep{2007CQGra..24.5985C}. Therefore, the subsequent analysis based on the cosmography method only uses data with redshift less than 1.0.

\section{Methodology} \label{sec4}
\subsection{MCMC method}
For SNe Ia, the best fitting results of cosmological parameters are achieved by minimizing the chi-square ($\chi^{2}$),
\begin{equation}
        \chi^{2}_{SN} = \Delta \mu \, \mathbf{C}^{-1}_\mathrm{stat+syst} \, \Delta \mu ^\mathrm{T}, 
        \label{chi}
\end{equation}
where $\Delta \mu$ is the difference between the observational distance modulus $\mu_\mathrm{obs}$ and the theoretical distance modulus $\mu_\mathrm{th}$:
\begin{equation}
        \Delta \mu = \mu_\mathrm{obs}(z_{i}) - \mu_\mathrm{th}(P_{i},z_{i}),
        \label{Dmu}
\end{equation}
here, $P_{i}$ represents the parameter to be fitted. For the flat $\Lambda$CDM model, $P_{i}$ represents $\Omega_{m}$ and $H_{0}$. For the cosmography method, it represents the kinematic parameters of the Universe, such as $H_{0}$, $q_{0}$, $j_{0}$, $s_{0}$, $l_{0}$. Parameter $\mathbf{C}_\mathrm{stat+sys}$ is the total covariance matrices which consists of the statistical ($\mathbf{C}_\mathrm{stat}$) and systematic covariance matrices ($\mathbf{C}_\mathrm{sys}$). Its form is as follows
\begin{equation}
    \mathbf{C}_\mathrm{stat+sys} = \mathbf{C}_\mathrm{stat} +  \mathbf{C}_\mathrm{sys}.
        \label{C}
\end{equation}
The used datasets, $\mu_\mathrm{obs}$ and $\mathbf{C}_\mathrm{stat+sys}$, are provided by \citet{2022ApJ...938..110B} and can be obtained online\footnote{https://github.com/PantheonPlusSH0ES/DataRelease}. 

For the CCs dataset, the constraints of cosmological parameters can be given by minimizing the corresponding $\chi^{2}_{H(z)}$ which are as follow 
\begin{equation}
	\chi^{2}_{H(z)} = \sum_{i=1}^{35}\frac{(H(z)_{obs}(z_{i}) - H(z)_{th}(P_{i},z_{i}))^{2}}{\sigma^{2}_{i}},
\label{chihz}
\end{equation}
where $\sigma_{i}$ is the observational uncertainties of Hubble parameter measurements. 

The combination of Equations (\ref{chi}) and (\ref{chihz}) can give the cosmological constraints of the joint dataset consists of SNe Ia and H(z) datasets. The corresponding $\chi^{2}_{total}$ is given by
\begin{equation}
\chi^{2}_{total} = \chi^{2}_{SN} + \chi^{2}_{H(z)}.
\label{chitotal}
\end{equation}

Minimizing Equation (\ref{chitotal}), we can obtain the best fitting results of cosmological parameters from the joint dataset. In this work, the minimization was performed employing a Bayesian Markov chain Monte Carlo \citep[MCMC;][]{2013PASP..125..306F} method with the $emcee$ package\footnote{https://emcee.readthedocs.io/en/stable/}. All the fittings in this paper were obtained adopting this python package. The MCMC samples were plotted utilizing the $getdist$ package \citep{2019arXiv191013970L}.

\subsection{Dipole fitting method}
Dipole fitting (DF) method was proposed to find the fine structure constant ($\Delta \alpha$/$\alpha$) dipole \citep{2012MNRAS.422.3370K}. This approach constructs a dipole+monopole model which constitutes the first two term of the spherical harmonic expansion. Instead of ($\Delta \alpha$/$\alpha$), which corresponds to fine structure constant deviations from its earth measured value, \citet{2012PhRvD..86h3517M} first applied this method to probe the dark energy dipole based on the distance modulus $\mu$. The basic equation is as follows
\begin{equation}
(\frac{\Delta \mu(z)}{\mu(z)_{th}}) \equiv \frac{\mu(z)_{th} - \mu(z)_{obs}} {\mu(z)_{th}} = A\cos{\theta}+B,
\label{ddd}
\end{equation}
where $\cos{\theta}$ is the angle with the dipole axis, $A$ is the dipole magnitude, and $B$ is the monopole magnitude. Afterwards, it has been widely used for probing the preferred direction of cosmic anisotropy \citep{2018ChPhC..42k5103C,2018PhRvD..97l3515D}. Equation (\ref{ddd}) can be simplified to the following form 
\begin{equation}
\tilde{\mu}_{th} = \mu_{th}\times(1+A\cos{\theta}+ B),
\label{dfmu}
\end{equation}
where $\tilde{\mu}_{th}$ is the modified theoretical distance modulus. The angle with the dipole axis, $\cos{\theta}$, is defined by 
\begin{equation}
\cos{\theta} = \hat{\textbf{n}}\cdot\hat{\textbf{p}},
\label{theta}
\end{equation}
here, $\hat{\textbf{n}}$ and $\hat{\textbf{p}}$ correspond to the
dipole direction and the unit vector pointing to the position of SN Ia, respectively. In the galactic coordinate, the form of $\hat{\textbf{n}}$ can be written as
\begin{equation}
\hat{\textbf{n}} = \cos{(b)}\cos{(l)}\hat{\textbf{i}}+\cos{(b)}\sin{(l)}\hat{\textbf{j}}+sin{(b)}\hat{\textbf{k}}.
\label{n}
\end{equation}
For any observational source whose location is $(l_{i},b_{i})$, $\hat{\textbf{p}}$ is given by
\begin{equation}
\hat{\textbf{p}_{i}} = \cos{(b_{i})}\cos{(l_{i})}\hat{\textbf{i}}+\cos{(b_{i})}\sin{(l_{i})}\hat{\textbf{j}}+sin{(b_{i})}\hat{\textbf{k}}.
\label{p}
\end{equation}
Replace $\mu_{th}$ in Equation (\ref{Dmu}) with Equation (\ref{dfmu}),
we can obtain the constraints considering the dipole+monopole (dm) correction. If the observations are more favorable to an isotropic universe, the constraints on the dipole magnitude $A$ and monopole magnitude $B$ will be closer to zero.

\begin{figure}[htp]
	\centering
	\includegraphics[width=0.37\textwidth]{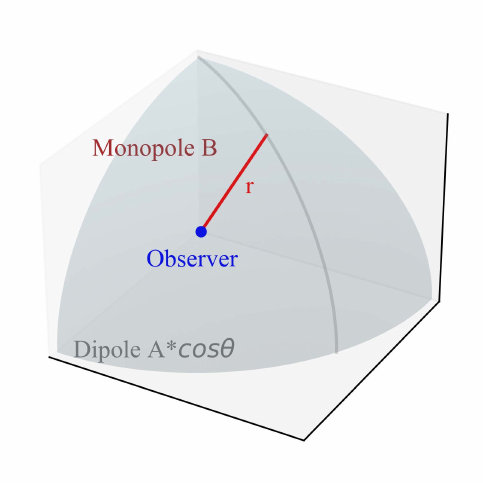}
	\caption{A schematic diagram for the dipole+monopole model. The blue dot represents the observer's position. The shape is a segment of a three-dimensional sphere with radius r.}
	\label{3d}       
\end{figure}

From Equation (\ref{dfmu}), it is easy to find that the parameter $B$ should be independent of direction and mainly characterizes whether the radial direction supports the cosmological principle. But the real observations come from different directions, and the anisotropy in different directions will also cause the $B$ value to deviate from zero. The correction term $A\cos{\theta}$ is a direction-dependent term, which mainly reflects whether different directions satisfy the cosmological principle. For a fixed redshift, it reflects the deviation of the spherical surface from the cosmological principle. In theory, the dipole is evaluated as a modulation around the monopole. Therefore, these two parameters can influence each other. In short, these two parameters reflect whether the observations support the cosmological principle in three-dimensional space. To help understanding, a schematic diagram is given in Figure \ref{3d}.  

It should be noted that for the constraints considering the dm correction, the dm correction is only applied to the Pantheon+ sample. The constraints of cosmological parameters (including $\Omega_{m}$, $H_{0}$, $q_{0}$, $j_{0}$, $s_{0}$, and $l_{0}$) are given by the joint sample which consists of SNe Ia and H(z). The ones of correction parameters ($l$, $b$, $A$, $B$) are given by the SNe Ia. 

\section{Results and discussions} \label{sec5}
\subsection{Cosmological constraints based on $\Lambda$CDM model}
\begin{figure}[htp]
	\centering
	\includegraphics[width=0.30\textwidth]{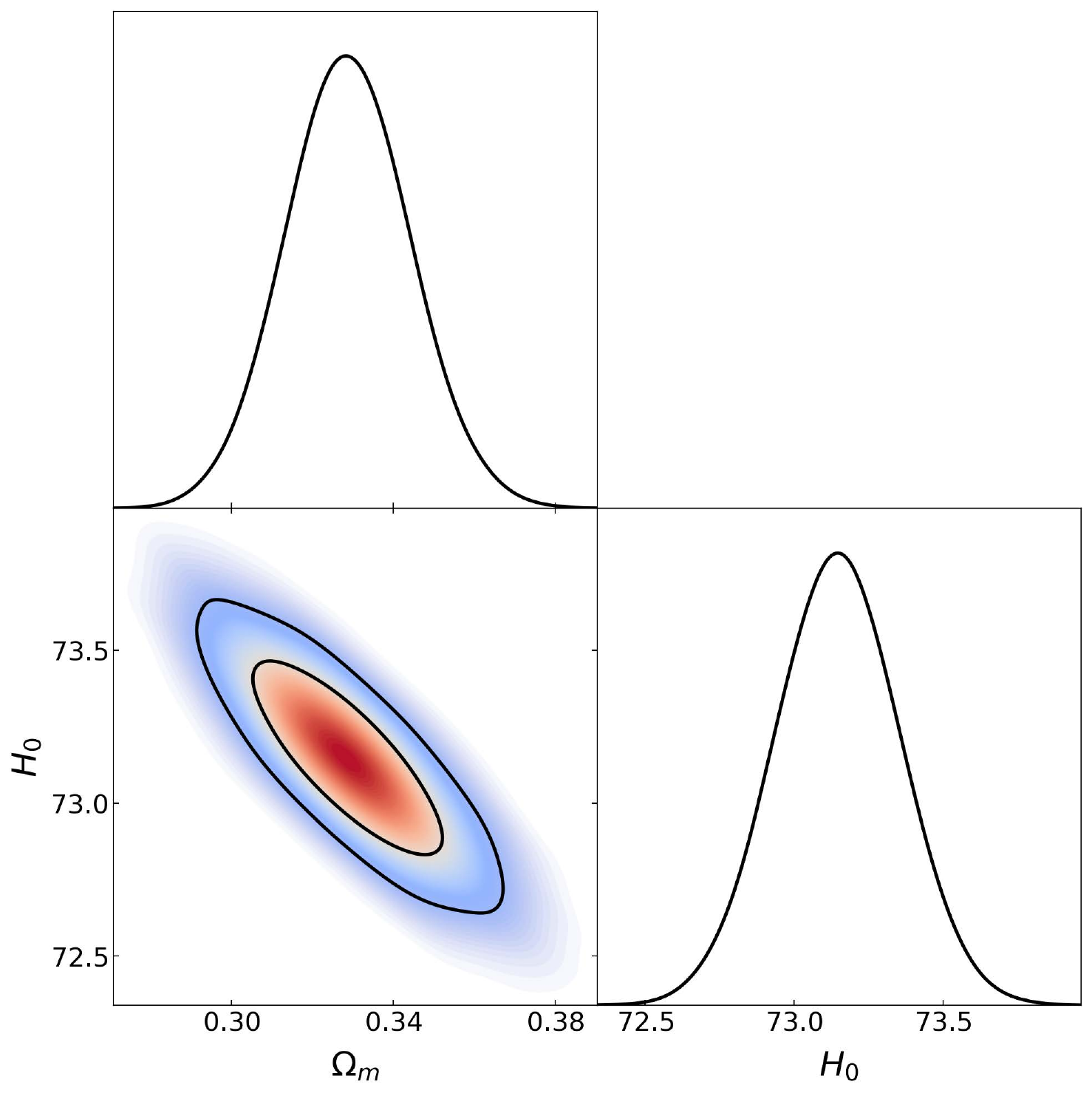}
	\caption{Confidence contours ($1\sigma$ and $2\sigma$) for the parameters space ($\Omega_{m}$ and $H_{0}$) from the joint dataset in the flat $\Lambda$CDM model. The fitting results are $\Omega_{m}$ = 0.33$\pm$0.02 and $H_{0}$ = 73.15$\pm$0.21 km/s/Mpc.}
	\label{snhz}       
\end{figure}

First, we gave the cosmological constraints of the $\Lambda$CDM model utilizing the joint dataset which consists of the Pantheon+ sample and 35 CCs, that is $\Omega_{m}$ = 0.33$\pm$0.02 and $H_{0}$ = 73.15$\pm$0.21 km/s/Mpc, as shown in Figure \ref{snhz}. The $\Omega_{m}$ result is consistent with the Planck 2018 result \citep{2020A&A...641A...1P} within 1$\sigma$ level and with DESI SN result \citep{2024arXiv240102929D} and DESI BAO result \citep{2024arXiv240403002D} within 2$\sigma$ level, but the $H_{0}$ result deviates significantly. The $H_{0}$ result is in line with that from late-time measurements within 1$\sigma$ range, including the local distance ladder \citep{2019ApJ...876...85R,2022ApJ...934L...7R}, Megamaser \citep{2020ApJ...891L...1P} and the H0LiCOW results \citep{2020MNRAS.498.1420W}. 
\begin{figure}[htp]
	\centering
	\includegraphics[width=0.44\textwidth]{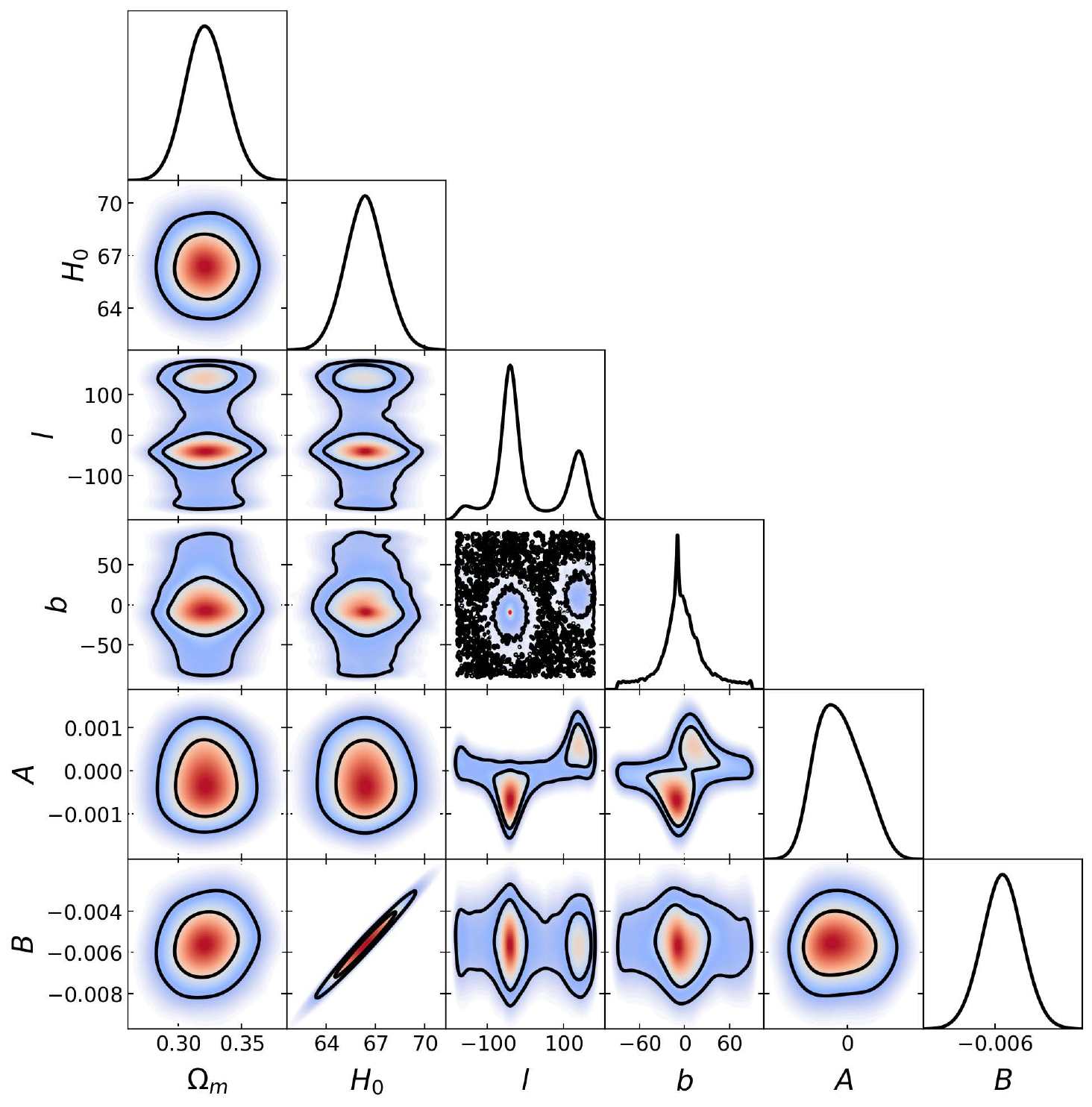}
	\caption{Confidence contours ($1\sigma$ and $2\sigma$) for the parameters space ($\Omega_{m}$, $H_{0}$, $l$, $b$, $A$ and $B$) from the joint dataset in the dipole+monopole model. The constraints of cosmological parameters are $\Omega_{m}$ = 0.32$^{+0.02}_{-0.02}$, $H_{0}$ = 66.38$^{+1.22}_{-1.22}$ km/s/Mpc. For correction parameters ($l$, $b$ and $A$), there are two constraints. The two constraints on parameter A are negative of each other, and the figure shows the superposition effect of two Gaussian distributions. The more accurate constraints of the correction parameters are given by Figure \ref{snhz_dipole_2}.}
	\label{snhz_dipole}   
\end{figure} 

After that, utilizing the DF method, we gave the constraints of the $\Lambda$CDM model with the dm correction, as shown in Figure \ref{snhz_dipole}. For convenience, we refer to this modified model as dm correction model. The constraints of cosmological parameters are $\Omega_{m}$ = 0.32$\pm$0.02, $H_{0}$ = 66.38$\pm$1.22 km/s/Mpc which are both agreement with the Planck 2018 results \citep[early-time measurements;][]{2020A&A...641A...1P} within 1$\sigma$ range. Combining the results from the $\Lambda$CDM model and the dm correction model, it can be found that $\Omega_{m}$ constraints are consistent with each other, but there are obvious differences in $H_{0}$. The dm correction reduces the $H_{0}$ constraints from 73.15 km/s/Mpc consistent with the late-time measurements to 66.38 km/s/Mpc which in line with the early-time ones. In other word, the dm correction model can effectively alleviate the $H_{0}$ tension. Next step, how to quantitatively describe the degree of relief for the Hubble tension will be discussed.

\begin{figure}[htp]
	\centering
	\includegraphics[width=0.32\textwidth]{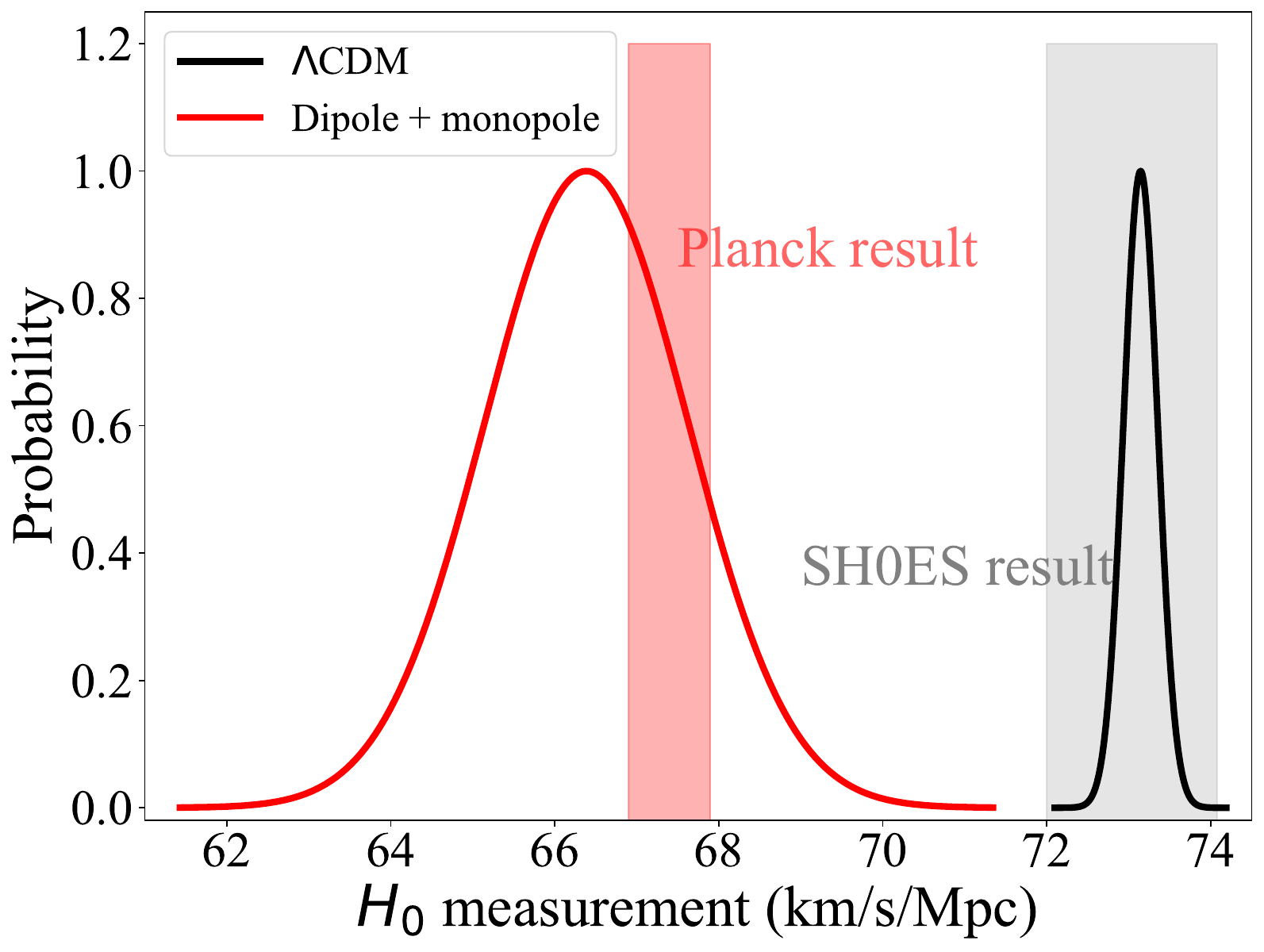}
	\caption{Comparison of our $H_{0}$ results with the Planck 2018 result \citep[67.36$\pm$0.50 km/s/Mpc;][]{2020A&A...641A...1P}and the SH0ES result \citep[73.04$\pm$1.04 km/s/Mpc;][]{2022ApJ...934L...7R}.}
    \label{com} 
\end{figure}

Until now, there is no unified approach to quantifying the alleviated degree of Hubble tension. Therefore, attempts to analyze the extent to which schemes can alleviate the Hubble tension from different perspectives should be encouraged. Here, we presented two kinds of schemes to quantify the alleviate degree, that is schemes (I) and (II). The scheme (I) is to compare our results with the early-time and late-time measurements respectively, and calculate the overall degree of agreement. The higher agreement degree, the higher the relieve degree. The percentage of reduction (\%Reduc= $C_{\rm late} \times C_{\rm early}$) is calculated by the following formula \citep{2022MNRAS.517..576H}:
\begin{eqnarray}
	C_{\rm late}(\sigma) &=& \frac{ H_{\rm 0, {\Lambda}CDM} - H_{\rm 0, SH0ES}}{\sqrt{ \sigma_{\rm {\Lambda}CDM}^{2}+\sigma_{\rm SH0ES}^{2}}}, 
	\label{xfit1}
\end{eqnarray}
\begin{eqnarray}
	C_{\rm early}(\sigma) &=& \frac{ H_{\rm 0, dipole} - H_{\rm 0, Planck}}{\sqrt{\sigma_{\rm dipole}^{2} + \sigma_{\rm Planck}^{2}}},
	\label{xfit}
\end{eqnarray}
where $C_{\rm late}$ is the agreement degree between the $H_{0}$ measurement of $\Lambda$CDM model ($H_{\rm 0, {\Lambda}CDM}$) and ones from the SH0ES collaboration ($H_{\rm 0, SH0ES}$), and $C_{\rm early}$ represents the the agreement degree between the $H_{0}$ measurement of dm correction model ($H_{\rm 0, dipole}$) and ones from the \emph{Planck} collaboration ($H_{\rm 0, Planck}$). Parameters $\sigma_{\rm {\Lambda}CDM}$, $\sigma_{\rm SH0ES}$, $\sigma_{\rm dipole}$ and $\sigma_{\rm Planck}$ are the corresponding 1$\sigma$ errors of $H_{0}$ constraints. Finally, the relieved degree of $H_{0}$ tension is estimated as 56\%. The scheme (II) is to utilize Equation (\ref{xfit}) to calculate the residual value of the $H_{0}$ tension between the early measurements and $H_{0}$ result which derived from the late-time measurements employing the dm correction model. The calculation results show that the residual tension is 0.77$\sigma$, and the corresponding alleviate degree is 73\%. For better understanding, we also put our results on the same canvas together with the early- and late-time $H_{0}$ measurements, as shown in Figure \ref{com}. From Figure \ref{com}, it is obvious that the Hubble tension has been effectively alleviated. 

It is worth to noting that the constraints of correction parameters ($l$, $b$ and $A$) give two set of best fitting results, that is, the correction parameters ($l$, $b$ and $A$) have two significance peaks, as shown in Figure \ref{snhz_dipole}. It is more easily seen from panels $l-b$, $l-A$, and $b-A$ of Figure \ref{snhz_dipole}. In order to obtain more precisely constraints on the correction parameters ($l$, $b$, $A$ and $B$), a suitable parameter space is necessary to make sure the direction parameters only with a single best fitting result. In short, Figure \ref{snhz_dipole} was divided into two parts by splitting the parameter space of $l$ and $b$. The corresponding results are shown in Figure \ref{snhz_dipole_2}. The left panel of Figure \ref{snhz_dipole_2} gives $\Omega_{m}$ = 0.32$\pm$0.02, $H_{0}$ = 66.42$^{+1.23}_{-1.19}$ km/s/Mpc, $l$ = 319.79$^{\circ}$$^{+24.86}_{-21.20}$, $b$ = $-$9.30$^{\circ}$$^{+16.70}_{-18.51}$, $A$ = $-$5.2$\pm$3.6$\times$10$^{-4}$, $B$ = $-$5.6$\pm$1.0$\times$10$^{-3}$. The right panel of Figure \ref{snhz_dipole_2} shows $\Omega_{m}$ = 0.32$\pm$0.02, $H_{0}$ = 66.39$\pm$1.22 km/s/Mpc, $l$ = 140.50$^{\circ}$$^{+24.96}_{-21.14}$, $b$ = 10.06$^{\circ}$$^{+19.67}_{-15.09}$, $A$ = 5.2$^{+3.4}_{-3.8}$$\times$10$^{-4}$, $B$ = $-$5.6$\pm$1.1$\times$10$^{-3}$. Combining the results of Figures \ref{snhz_dipole} and \ref{snhz_dipole_2}, we find that the constraints of cosmological parameters ($\Omega_{m}$ and $H_{0}$) only have minor differences. Therefore, the best fitting results of cosmological parameters are given by the total parameter space (Figure \ref{snhz_dipole}), and the constraints of correction parameters are given by the truncate parameter space (Figure \ref{snhz_dipole_2}). In addition, these two preferred directions of cosmic anisotropy are symmetrical in the galactic coordinate system, and the absolute values of the corresponding dipole magnitude $A$ are nearly equal. 

Utilizing the Pantheon+ sample and the region fitting method, \citet{2024A&A...681A..88H} found a local void region ($313.43_{-18.17}^{+19.64}$, $-16.81_{-10.69}^{+11.08}$) and a preferred direction of the cosmic anisotropy ($308.39_{-48.66}^{+47.60}$, $-18.19_{-28.79}^{+21.05}$). Our preferred directions are consistent with their results. Combining our results with their all-sky distributions of cosmological parameters (Figure 3), we can find that the direction (319.79$^{\circ}$$^{+24.86}_{-21.20}$, -9.30$^{\circ}$$^{+16.70}_{-18.51}$) is consistent with the region of lower matter density and higher expansion rate of the universe. The direction (140.50$^{\circ}$$^{+24.96}_{-21.14}$, 10.06$^{\circ}$$^{+19.67}_{-15.09}$) is in agreement with the region of higher matter density and lower expansion rate of the universe. Moreover, the preferred directions of cosmic anisotropy we found are consistent with those obtained from other researches within 1$\sigma$, such as Union2 \citep{2012PhRvD..86h3517M,2015MNRAS.446.2952C}, Union2.1 \citep{2019EPJC...79..783S}, Keck+VLT \citep{2012PhRvD..86h3517M}, JLA \citep{2016MNRAS.456.1881L}, Pantheon \citep{2018MNRAS.478.5153S,2019MNRAS.486.5679Z}, SN-Q \citep{2020A&A...643A..93H}, $\Delta \alpha$ \citep{2011PhRvL.107s1101W,2012MNRAS.422.3370K}, Galaxy cluster \citep{2020A&A...636A..15M,2021A&A...649A.151M}, Infrared galaxies \citep{2014MNRAS.445L..60Y,2017MNRAS.464..768B}, Dark flow \citep{2010ApJ...712L..81K}, Bulk flow \citep{2023MNRAS.524.1885W}, Quasar \citep{2021EPJC...81..948Z}, Pantheon+ \citep{2024arXiv240614827H} and so on.

We also carried out a redshift tomography analysis to discuss the redshift dependence of our findings, and a short discussion around the influence of parameters $A$ and $B$ on the Hubble constant. Through increasing cut-off redshift, we gave the corresponding constraints of the $\Lambda$CDM model and the dm correction model, respectively. There are total six groups. Detailed information is shown in Table \ref{tab:fit}, including the number of used sources and the constraints from the $\Lambda$CDM model and the dm correction model. Looking around the best fitting results of different cut-off redshifts, it is easy to find that the $H_{0}$ constraints of $\Lambda$CDM model are near by 72.70 km/s/Mpc, which is biased towards the distance ladder result \citep{2022ApJ...934L...7R}. The ones from the dm correction model are relatively small, which is biased towards the Planck 2018 result \citep{2020A&A...641A...1P}. For the correction parameters, the preferred directions from different cutoff redshifts are consistent with each other. The dipole magnitude $A$ and monopole magnitude $B$ decrease slightly with redshift, and $A$ has a minimum value at cutoff redshift = 1.00. The values of $\Omega_{m}$ also decrease with redshift, and $\Omega_{m}$ of the dm correction model has a minimum value at cutoff redshift = 0.50. All in all, at different cutoff redshift, considering the dm correction can both obviously reduce the late-time $H_{0}$ value, that effectively alleviates the Hubble tension. For discussing the effect of parameters $A$ and $B$ on the Hubble constant, we consider fixing $A$ = 0 and repeat the analysis. The final results we get are $\Omega_{m}$ = 0.32$\pm$0.02, $H_{0}$ = 66.40$\pm1.20$ km/s/Mpc, $B$ = $-$5.5$\pm$1.0$\times$$10^{-3}$. These results are in line with the previous ones. The $B$ value increases slightly, $\Delta B$ equal to 1.0$\times$$10^{-4}$. From Equation (\ref{dfmu}), we can see that when parameter $A$ is fixed at 0, the effect of the correction term $A\cos{\theta}$ on $\mu_{th}$ might be partially compensated by the value of $\Delta B$. A small $\Delta B$ shows that the decrease in $H_{0}$ might be mainly affected by the monopole parameter $B$. It is worth noting that when we only consider the monopole $B$, the research scale will be reduced from three dimensions to one dimension, but the real data is three-dimensional. And the dipole anisotropy will make the one-dimensional test that ignores direction deviate from isotropy. Thus, the current study could not separate the effects of dipole $A$ and monopole $B$ on the Hubble constant. In order to distinguish the influence of the dipole $A$ and monopole $B$ on cosmological parameters, the most basic prerequisite is to have enough observational data in any direction and distributed at different redshifts. Obviously, the current SNe Ia observations do not meet such conditions.

All in all, non-zero $A$ and $B$ values violate the cosmological principle, leading to a small $H_{0}$ value. If the violation of the cosmological principle is due to a physical mechanism and not to systematic or statistical fluctuations, it would support the claim that the Hubble tension stems from new physics beyond the standard $\Lambda$CDM model. There are many possible physical reasons why $A$ and $B$ are not zero. It might be caused by a single factor, for example an off-center observer in a bubble of distinct physics or systematics \citep{2023PhRvD.108f3509P}. Of course, they might also be caused by a combination of factors. For instant, the factors that cause monopole $B$ to be non-zero might be an area of low matter density in a single direction, while a non-zero dipole $A$ might be caused by an inhomogeneous scalar field which couples to electromagnetism through a non-minimal coupling and whose potential energy could provide the dark energy required for accelerating expansion \citep{2012PhRvD..86h3517M}. \citet{Alnes:2006pf} and \citet{Grande:2011hm} have discussed the dipole nature of observations of off-center observers located in spherically symmetric inhomogeneities. Finally, it hints that the Hubble tension might be also caused by a single factor or a combination of factors. 

The current results are obtained according to the $\Lambda$CDM model. In order to check the model dependence of results, we replace the $\Lambda$CDM model with the cosmography method to make a cross check. More detailed information are shown in next subsection.

\subsection{Cross check by the cosmography method}
According to the characteristics of the cosmography method ($z$-redshift series), we choose to utilize the part with the joint dataset redshift less than 1.0 to make the cross check. At this time, the low-redshift joint dataset contains 1676 SNe Ia and 28 CCs. For convenience, we refer to this sample as LJ+ sample. Afterwards, based on the Akaike information criterion (AIC) and the Bayesian information criterion (BIC), we compared the performance of different expansion orders (including the second-order, third-order and forth-order) to find the most suitable expansion order. Out of necessity, we briefly introduce the model comparison methods used. We note that AIC and BIC are the last set of techniques that can be employed for model comparison based on information theory, and they are widely used in cosmological model comparisons \citep{2020ApJ...888...99W,2023PhRvD.108h3024H,2024A&A...689A.165L}. AIC and BIC are defined by \citet{1974ITAC...19..716A} and \citet{1978AnSta...6..461S}, respectively. The corresponding definitions read as follows: 
\begin{eqnarray}
        AIC &=& 2n + \chi_{min}^{2},\\
        BIC &=& n\log{N} + \chi_{min}^{2},
        \label{eq:AIC}
\end{eqnarray}
where $N$ is the total number of data points, $n$ is the number of free parameters, and $\chi_{min}^{2}$ is the value of $\chi^{2}$ calculated with the best fitting results. The model that has lower values of AIC and BIC will be the suitable model for the used dataset. Moreover, we also calculated the differences between $\Delta$AIC and $\Delta$BIC with respect to the corresponding flat $\Lambda$CDM values to measure the amount of information lost by adding extra parameters in the statistical fitting. The values of $\Delta$AIC and $\Delta$BIC exceeding the threshold (8-12) indicate that the model under investigation performs better than the reference model \citep{doi:10.1080Kass}.

\begin{figure}[htp]
	\centering
	\includegraphics[width=0.23\textwidth]{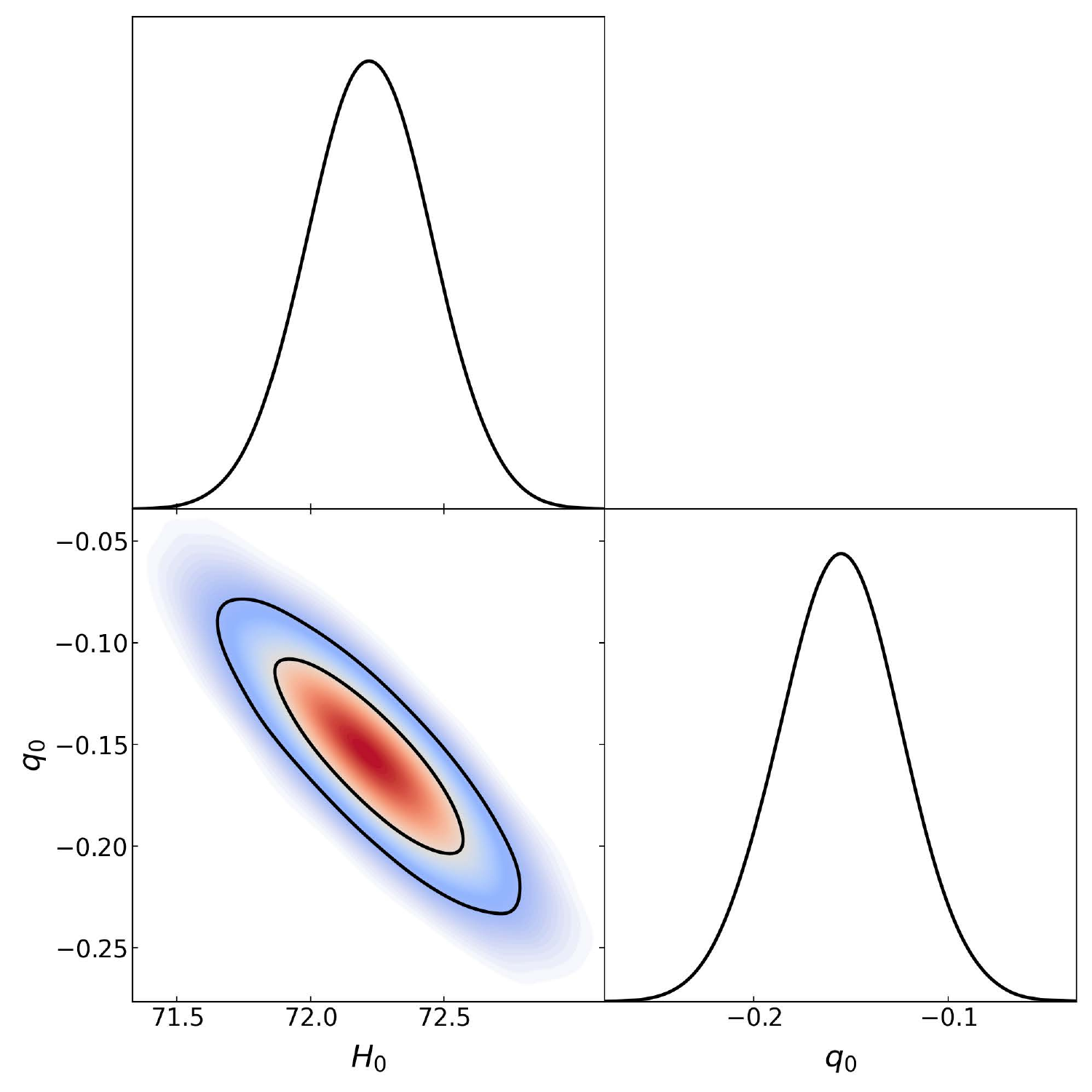}
	\includegraphics[width=0.23\textwidth]{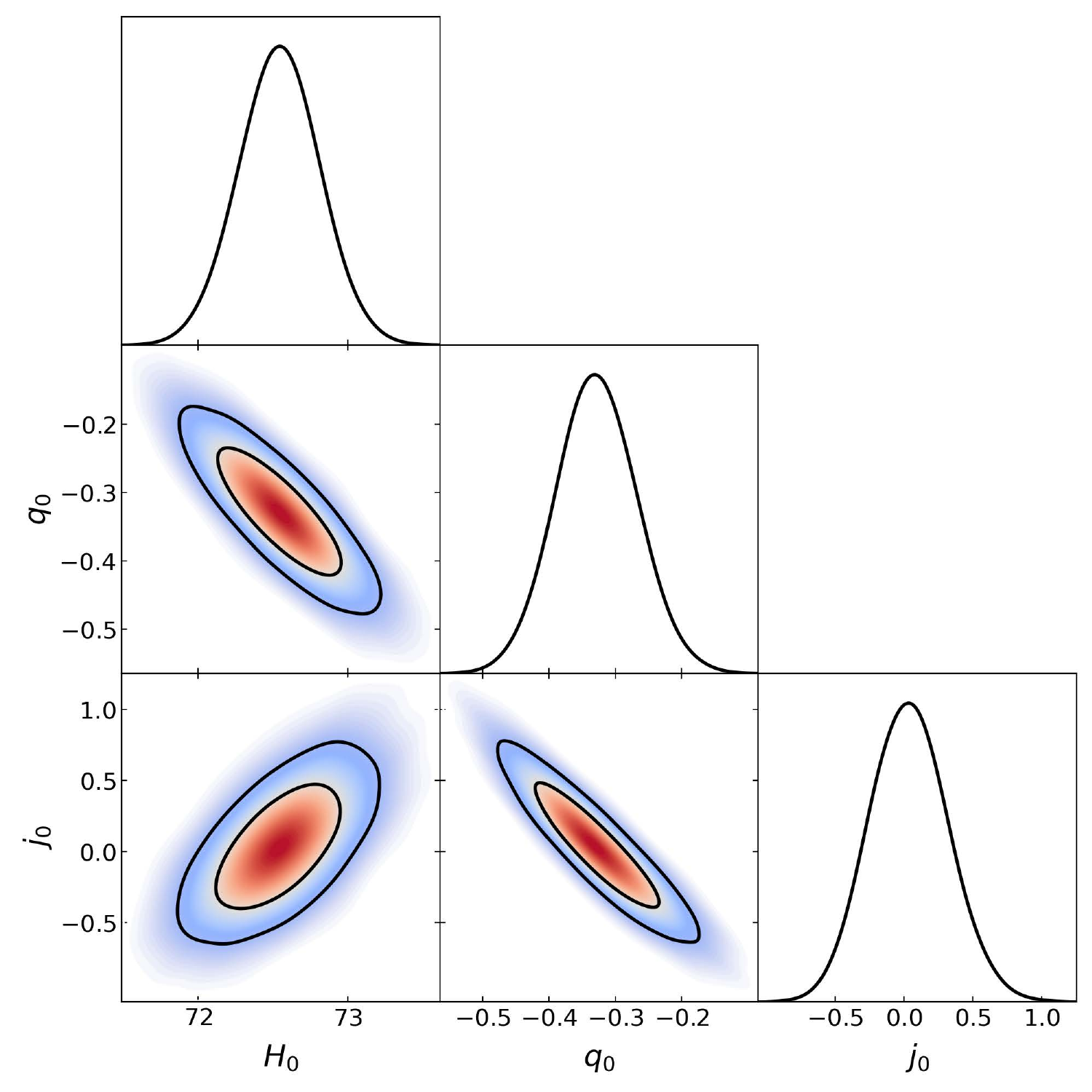} \\
	\includegraphics[width=0.25\textwidth]{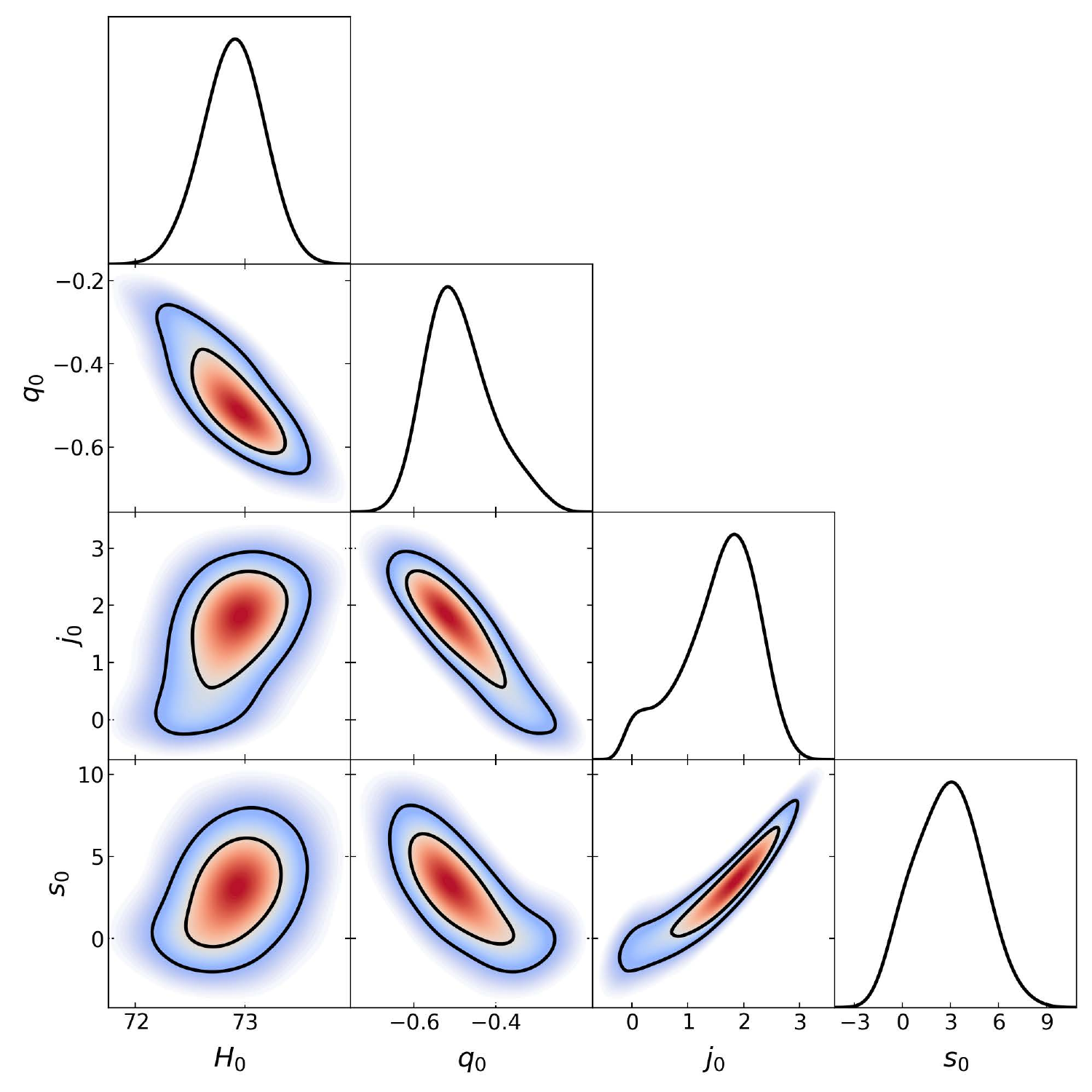}
	\caption{Confidence contours ($1\sigma$ and $2\sigma$) for the parameters space ($H_{0}$, $q_{0}$, $j_{0}$ and $s_{0}$) from the low redshift joint dataset utilizing the cosmography method. }
	\label{Taylor_select}       
\end{figure}

\begin{figure}[htp]
	\centering
	\includegraphics[width=0.42\textwidth]{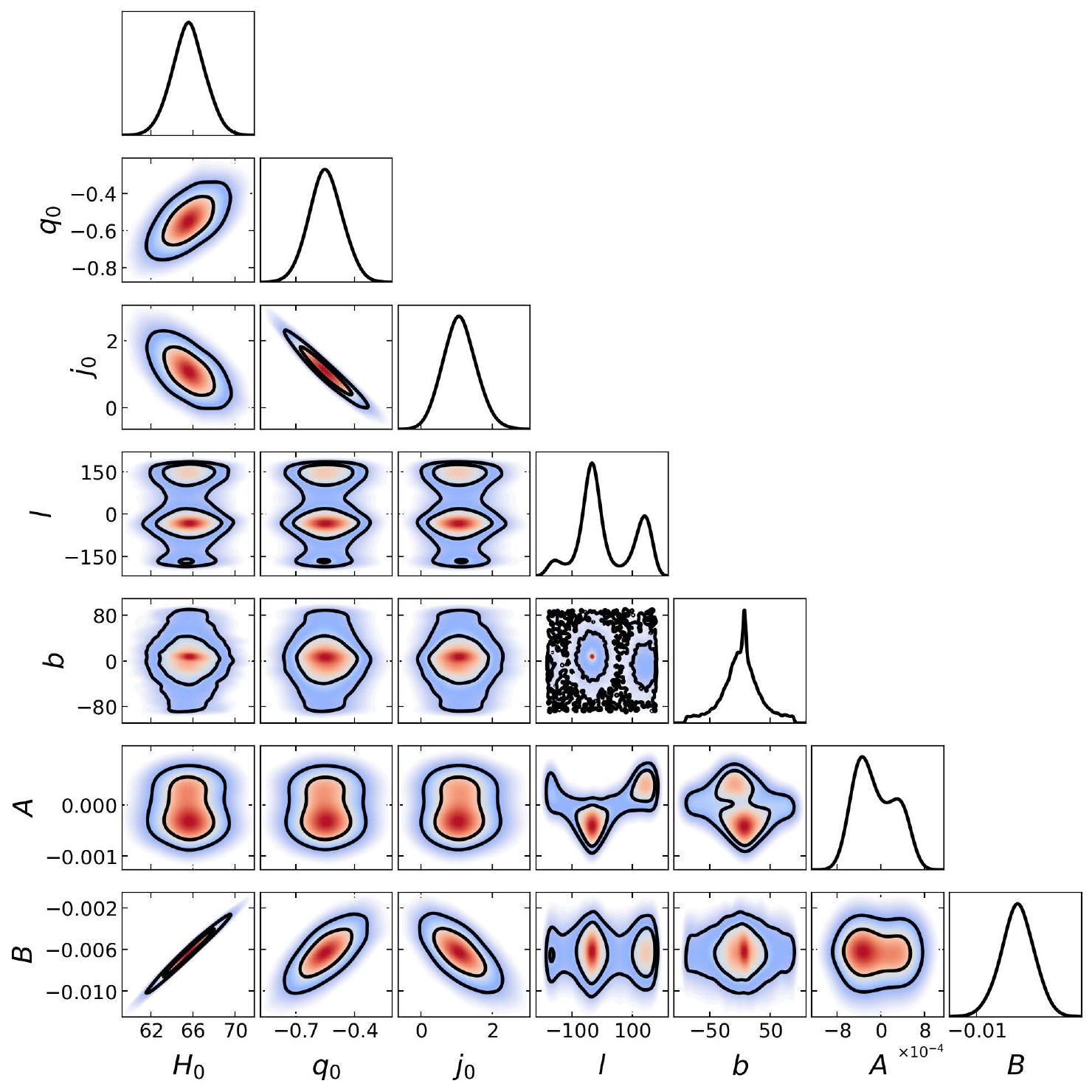}
	\caption{Confidence contours ($1\sigma$ and $2\sigma$) for the parameters space ($H_{0}$, $q_{0}$, $j_{0}$, $l$, $b$, $A$ and $B$) from the low redshift joint dataset using the third-order expansion. The constraints of cosmological parameters are $H_{0}$ = 72.54$\pm$0.27 km/s/Mpc, $q_{0}$ = $-$0.33$\pm$0.06 and $j_{0}$ = 0.03$^{+0.29}_{-0.28}$. For correction parameters ($l$, $b$ and $A$), there are two constraints. The two constraints on parameter A are negative of each other, and the figure shows the superposition effect of two Gaussian distributions. The more accurate constraints of the correction parameters are given by Figure \ref{z_dipole_2}.}
	\label{z_dipole}       
\end{figure}

The cosmographic constraints and corresponding statistical information are summarized in Table \ref{tab:taylor}. The statistical results show that the LJ+ sample favors the third-order expansion, that is $\Delta$AIC = $-$8.53 and $\Delta$BIC = $-$3.09. However, $\Delta$BIC = $-$3.09 shows that this preference is not statistically significant. The corresponding fitting results are $H_{0}$ = 72.54$\pm$0.27 km/s/Mpc, $q_{0}$ = $-$0.33$\pm$0.06 and $j_{0}$ = 0.03$^{+0.29}_{-0.28}$. The model-independent $H_{0}$ constraints are consistent with the previous results of the flat $\Lambda$CDM model. In terms of the comparison results of different expansion orders, we decided to employ the third-order expansion. The third-order expansion was combined with the DF method to obtain the dipole-monopole corrected constraints, as shown in Figures \ref{z_dipole} and \ref{z_dipole_2}. According to previous researches based on the $\Lambda$CDM model, the constraints on the cosmological parameters and preferred directions come from different parameter spaces respectively. The constraints of cosmological parameters are obtained from the total parameter space, that is, $H_{0}$ = 65.59$\pm$1.59 km/s/Mpc, $q_{0}$ = $-$0.55$^{+0.09}_{-0.08}$ and $j_{0}$ = 1.07$^{+0.48}_{-0.46}$, as shown in Figure \ref{z_dipole}. The $H_{0}$ constraint is in line with the previous correction $H_{0}$ based on the $\Lambda$CDM model. The constraints for the two sets of correction parameters are shown in Figure \ref{z_dipole_2}, that is, ($l$ = 326.85$^{+30.24}_{-30.78}$, $b$ = 7.52$^{+24.38}_{-26.65}$, $A$ = $-$3.2$\pm$2.2$\times 10^{-4}$, $B$ = $-$6.3$\pm$1.5$\times 10^{-3}$) and ($l$ = 140.50$^{+24.96}_{-21.14}$, $b$ = 10.06$^{+19.67}_{-15.09}$, $A$ = 5.2$^{+3.4}_{-3.8}$$\times 10^{-4}$, $B$ = $-$5.6$\pm$1.1$\times 10^{-3}$). According to Equations (\ref{xfit1}) and (\ref{xfit}), the relieve degree of $H_{0}$ tension by the schemes (I) and (II) are 42\% and 64\%, respectively. Obviously, the reanalysis based on the cosmography method also suggests that the dipole-monopole correction can effectively reduce the $H_{0}$ constraint.  

\subsection{Comparison with the BAO measurements}
\begin{figure}[htp]
	\centering
	\includegraphics[width=0.35\textwidth]{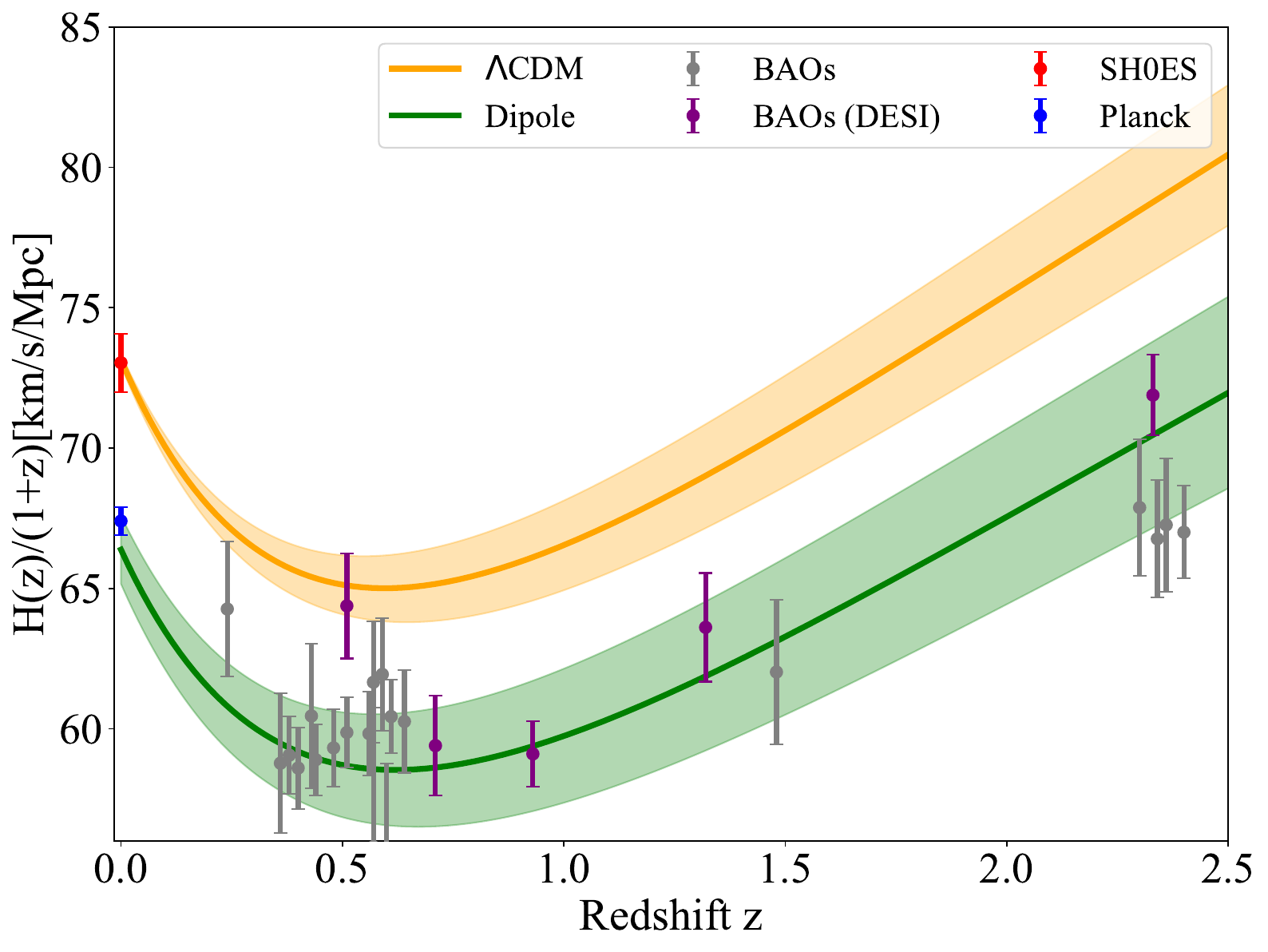}
        \caption{Variation of $H(z)/(1+z)$ with redshift $z$. Orange solid line and shaded region represent the theoretical prediction and the corresponding 1$\sigma$ range based on the constraints of $\Lambda$CDM model, respectively. Green represents the result of the $\Lambda$CDM model considering the dipole correction. Purple points represent DESI BAO measurements, and the gray points represent other BAO measurements.}
	\label{com_HZ}       
\end{figure}

Based on the constraints of the $\Lambda$CDM model and the dm correction model, we plotted the variation of $H(z)/(1+z)$ with redshift $z$ to intuitively display the difference which caused by the dm correction, as shown in Figure \ref{com_HZ}. The theoretical predictions of $H(z)$ are given by Equation (\ref{hz}). In Figure \ref{com_HZ}, we also marked the BAO measurements including the DESI BAO measurements. Not all BAO measurements are considered. We selected some BAO measurements with smaller errors based on a simple screening criterion which requires the error less than 5.0 for redshifts below 0.50 and less than 10.0 for redshifts above 0.50. There is total 25 BAOs measurements consist of 5 DESI BAOs and 20 other BAOs. More detail information about BAOs are shown in Table \ref{tab:BAO}.

\begin{table}\footnotesize
	\caption{BAO measurements including 5 DESI BAOs (in units of $\textrm{km}~\textrm{s}^{-1} \textrm{Mpc}^{-1}$). \label{tab:BAO}}
	\centering
	\begin{tabular}{cccc}
		\hline\hline
   No.	&	$z$  & $H(z)$  & Reference  \\  
  \hline
  \textbf{DESI:}	&   &   &   \\
  (1)	&	$0.51$    & $97.21\pm2.83$ &  \cite{2024arXiv240403002D} \\
  (2)	&	$0.71$    & $101.57\pm3.04$ &  \cite{2024arXiv240403002D} \\
  (3)	&	$0.93$    & $114.07\pm2.24$ &  \cite{2024arXiv240403002D} \\
  (4)	&	$1.32$    & $147.58\pm4.49$ &  \cite{2024arXiv240403002D} \\
  (5)	&	$2.33$    & $239.38\pm4.80$ &  \cite{2024arXiv240403002D} \\
  \hline \hline
  \textbf{Other:}	&   &   &   \\
  (1) 	&	$0.24$    & $79.69\pm2.99$ & \cite{2009MNRAS.399.1663G} \\
  (2) 	&	$0.36$    & $79.93\pm3.39$ & \cite{2017MNRAS.469.3762W} \\
  (3) 	&	$0.38$    & $81.50\pm1.90$ & \cite{2017MNRAS.470.2617A} \\
  (4)     &	$0.40$    & $82.04\pm2.03$ &  \cite{2017MNRAS.469.3762W} \\
  (5) 	&	$0.43$    & $86.45\pm3.68$ &  \cite{2009MNRAS.399.1663G} \\
  (6) 	&	$0.44$    & $84.81\pm1.83$ &  \cite{2017MNRAS.469.3762W} \\
  (7) 	&	$0.48$    & $87.79\pm2.03$ & \cite{2017MNRAS.469.3762W} \\
  (8)  	&	$0.51$    & $90.40\pm1.90$ & \cite{2017MNRAS.470.2617A} \\
  (9) 	&	$0.56$    & $93.33\pm2.32$ & \cite{2017MNRAS.469.3762W} \\
  (10)	&	$0.57$    & $87.60\pm7.80$ & \cite{2013MNRAS.433.3559C} \\
  (11)	&	$0.57$    & $96.80\pm3.40$ & \cite{2014MNRAS.441...24A} \\
  (12)	&	$0.59$    & $98.48\pm3.19$ & \cite{2017MNRAS.469.3762W} \\
  (13)	&	$0.60$    & $87.90\pm6.10$ & \cite{2012MNRAS.425..405B} \\
  (14)	&	$0.61$    & $97.30\pm2.10$ & \cite{2017MNRAS.470.2617A} \\
  (15)	&	$0.64$    & $98.82\pm2.99$ &  \cite{2017MNRAS.469.3762W} \\
  (16)	&	$1.48$    & $153.81\pm6.39$ &  \cite{2020MNRAS.499..210N} \\
  (17)	&	$2.30$    & $224.0\pm8.0$ &  \cite{2013AA...552A..96B} \\
  (18)	&	$2.34$    & $223.0\pm7.0$ &  \cite{2015AA...574A..59D} \\
  (19)	&	$2.36$    & $226.0\pm8.0$ &  \cite{2014JCAP...05..027F} \\
  (20)	&	$2.40$    & $227.8\pm5.61$ &  \cite{2017AA...608A.130D} \\
  \hline\hline
	\end{tabular}
\end{table}

From Figure \ref{com_HZ}, it is easy to find that the dm correction significantly shifts the theoretical predictions of $H(z)/1+z$ downward. The BAO measurements are in line with the theoretical prediction reconstructed by the dm correction constraints within 1$\sigma$ range except datapoint at z = 0.51. This odd behavior gains our attention. There seems to be magic around redshift 0.50. \citet{2024arXiv240408633C} demonstrated that a $\sim$ 2$\sigma$ discrepancy with the Planck-$\Lambda$CDM cosmology in DESI Luminous Red Galaxy (LRG) data at $z_{eff}$ = 0.51 translates into an unexpectedly large $\Omega_{m}$ value, $\Omega_{m}$ = 0.67$^{+0.18}_{-0.17}$. They also confirmed that this anomaly drives the preference for $w_{0}$ $>$ $-$1 in DESI data confronting to the $w_{0}$$w_{a}$CDM model. Afterwards, the DESI collaboration confirmed that the data point at redshift z = 0.51 mildly offset (at the 2$\sigma$ significance level) from the $\Lambda$CDM expectation. In addition, there has been many researches pointed out that the transition redshift is $z_{t}$ $\sim$ 0.50, which is the position where the expansion of the universe changes from deceleration to acceleration \citep{2007ApJ...659...98R,2016JCAP...05..014M,2020JCAP...04..053J,2022EPJC...82.1165S,2023IJMPD..3250039K}. Utilizing the H(z) data with the Gaussian process (GP) method \citep{2011JMLR...12.2825P}, a late-time transition of $H_{0}$ occurs near the redshift of 0.50, that is, $H_{0}$ changes from a low value to a high one from early to late cosmic time \citep{2022MNRAS.517..576H}. This transition behavior can effectively reduces the Hubble tension by 70 per cent. In addition, \citet{2024A&A...681A..88H} tested the cosmological principle using the Pantheon+ sample and the region-fitting method. They found that there is an anisotropic minimum at redshift 0.50 from the redshift tomography analysis. Of course, these could be a coincidence, but it could also be a signal that hints at the existence of new physics beyond the standard cosmological model. 

\section{Summary} \label{sec6}
In this paper, we investigated the impact of dm correction on the $H_{0}$ constraints utilizing the joint dataset which consists of the Pantheon+ sample and latest CCs observations. After that, we also compared the final results with BAOs observations. From the investigation, we find that the dm correction can reduce the $H_{0}$ constraint from a larger value consistent with the SH0ES results \citep{2022ApJ...934L...7R} to a smaller value consistent with the Planck results \citep{2020A&A...641A...6P}. This finding can effectively alleviate the Hubble tension by up to 73\%. A redshift tomography analysis confirms this finding at different redshift ranges, and shows that correction parameter $|A|$ has a minimum value near $z$ = 1.00, and parameter $|B|$ decreases with increasing redshift. The behaviors of correction parameters hints that the cosmic anisotropy might originate from a low-redshift Universe. We also made a brief discussion around the effect of parameters $A$ and $B$ on the Hubble constant, and found that the decline in $H_{0}$ might be mainly contributed by the monopole $B$. However, since the real data comes from different directions, the impact of dipole $A$ on $H_{0}$ will be superimposed on the original $B$. Even if $A$ = 0 is fixed, it is impossible to completely eliminate the influence of the spatial position of the observations on $B$. The current data conditions cannot strictly distinguish the effects of the dipole $A$ and the monopole $B$ on the Hubble constant. In addition, we also designed model-independent reanalyses which also support our previous finding from the $\Lambda$CDM model. Finally, BAOs measurements including 5 DESI BAOs are used for comparison with the final results. All the BAOs measurements are in line with the theoretical prediction of $\Lambda$CDM model considering the dm correction within 1$\sigma$ range except datapoint at $z$ = 0.51. Many studies have shown that redshifts around 0.50 seem a bit unusual \citep{2020JCAP...04..053J,2022EPJC...82.1165S,2023IJMPD..3250039K,2022MNRAS.517..576H,2024arXiv240408633C,2024arXiv240403002D,2024A&A...681A..88H}. 

In one word, from the Pantheon+ sample and the latest CCs dataset we find the violation of cosmological principle on a $10^{-4}$ level dipole and a $10^{-3}$ level monopole, which could effectively alleviate the Hubble tension. Our research hints that the Hubble tension might be related to the violation of the cosmological principle. This possibility must be taken seriously. The specific physical reasons for the violation of the cosmological principle need further targeted research.

\begin{acknowledgments}
Anonymous reviewer provided many constructive comments, for which we are deeply grateful. This work was supported by the National Natural Science Foundation of China (grant No. 12273009) and the China Manned Spaced Project (CMS-CSST-2021-A12). 
\end{acknowledgments}


\bibliographystyle{aasjournal}

\appendix
\section{Additional materials}
\begin{table}[h]
\setlength{\tabcolsep}{2.0pt}
\renewcommand{\arraystretch}{1.5}
\caption{Constraints of the $\Lambda$CDM model and the dipole correction model from the Pantheon+ sample and the latest H(z) dataset. \label{tab:fit}}
	\centering
	\begin{tabular}{ccc|cc|cccccc}
		\hline\hline
		$z_{max}$ & SNe Ia & H(z)& $\Omega_{m}^{*}$  & $H_{0}^{*}$  &  $\Omega_{m}^{+}$  & $H_{0}^{+}$  & l$^{+}$ & b$^{+}$ & $A^{+}$ & $B^{+}$ \\ 
		& Num.  &  Num.	&   & $\textrm{km}~\textrm{s}^{-1} \textrm{Mpc}^{-1}$  &    & $\textrm{km}~\textrm{s}^{-1} \textrm{Mpc}^{-1}$  & [$^{\circ}$] & [$^{\circ}$] & $\times$10$^{-4}$ & $\times$10$^{-3}$ \\
		\hline
		0.20  & 948 & 7	& 0.46$\pm$0.06 & 72.40$\pm$0.33 & 0.39$^{+0.07}_{-0.06}$  & 64.00$^{+2.06}_{-1.97}$ & 339.22$^{+18.83}_{-19.79}$ & $-$5.40$^{+14.58}_{-15.47}$ & $-$7.6$^{+2.7}_{-3.0}$ & $-$7.5$\pm$1.9 \\
		   &  &	&  &  &    &   & 159.25$^{+18.03}_{-19.54}$ & 5.66$^{+15.42}_{-14.46}$ & 7.6$\pm$2.9 & $-$7.4$\pm$1.9 \\		
    	0.30  & 1200 & 9 & 0.44$\pm$0.04 & 72.47$\pm$0.30 & 0.36$\pm$0.04  & 64.21$^{+1.94}_{-1.90}$ & 336.52$^{+14.80}_{-14.85}$ & 2.89$^{+11.87}_{-12.29}$ & $-$7.8$\pm$2.3 & $-$7.3$\pm$1.9 \\
    	   &  &	&  &  &    &   & 156.54$^{+14.81}_{-15.26}$ & $-$2.96$^{+11.87}_{-11.87}$ & 7.7$\pm$2.3 & $-$7.3$\pm$1.8 \\
    	0.40  & 1393 & 13 & 0.37$\pm$0.03 & 72.74$^{+0.27}_{-0.26}$ & 0.31$\pm$0.03  & 64.67$^{+1.79}_{-1.85}$ & 329.46$^{+19.54}_{-19.19}$ & 10.15$^{+15.46}_{-14.90}$ & $-$5.7$^{+2.0}_{-2.3}$ & $-$7.1$\pm$1.7 \\
    	   &  &	&  &  &    &   & 149.27$^{+18.61}_{-19.45}$ & $-$10.37$^{+14.55}_{-15.89}$ & 5.8$\pm$2.2 & $-$7.1$\pm$1.7 \\
    	0.50  & 1491 & 18 & 0.35$\pm$0.02 & 72.86$\pm$0.25 & 0.30$\pm$0.03  & 65.08$^{+1.64}_{-1.60}$ & 336.02$^{+25.34}_{-23.99}$ & 14.34$^{+20.34}_{-18.73}$ & $-$4.4$^{+2.0}_{-2.4}$ & $-$6.8$\pm$1.5 \\
    	   &  &	&  &  &    &   & 155.55$^{+22.10}_{-23.50}$ & $-$14.54$^{+18.33}_{-20.13}$ & 4.5$\pm$2.1 & $-$6.8$\pm$1.5 \\
    	1.00  & 1676 & 28 & 0.35$\pm$0.02 & 72.96$\pm$0.22 & 0.33$\pm$0.02  & 66.00$^{+1.27}_{-1.29}$ & 326.22$^{+29.18}_{-33.74}$ & 5.83$^{+25.12}_{-28.83}$ & $-$3.1$\pm$2.3 & $-$5.8$\pm$1.1 \\
    	   &  &	&  &  &    &   & 146.88$^{+29.78}_{-29.20}$ & $-$6.43$^{+26.33}_{-25.34}$ & 3.3$\pm$2.2 & $-$5.8$\pm$1.1 \\
    	2.26  & 1701 & 35 & 0.33$\pm$0.02 & 73.15$\pm$0.21 & 0.32$\pm$0.02  & 66.38$\pm$1.22 & 319.79$^{+24.86}_{-21.20}$ & $-$9.30$^{+16.70}_{-18.51}$ & $-$5.2$\pm$3.6 & $-$5.6$\pm$1.0 \\
    	   &  &	&  &  &    &   & 140.50$^{+24.96}_{-21.14}$ & 10.06$^{+19.67}_{-15.09}$ & 5.2$^{+3.4}_{-3.8}$ & $-$5.6$\pm$1.1 \\
		\hline\hline
	\end{tabular}
    \begin{itemize}	
	    \footnotesize
	    \item[*] $\Lambda$CDM model.
	    \item[+] Dipole model.
    \end{itemize}
\end{table}

\begin{table}[h]
\setlength{\tabcolsep}{2.0pt}
\renewcommand{\arraystretch}{1.5}
	\caption{Detailed information of cosmographic constraints. \label{tab:taylor}}
	\centering
	\begin{tabular}{cccccc|ccccc}
		\hline\hline
		Model & $\Omega_{m}$ & $H_{0}$  & $q_{0}$  & $j_{0}$ & $s_{0}$ & $\chi^{2}$  &  AIC & BIC & $\Delta$AIC & $\Delta$BIC    \\
		 &   & $\textrm{km}~\textrm{s}^{-1} \textrm{Mpc}^{-1}$  &   &   &   &   &  &  &  & \\
		 \hline
		$\Lambda$CDM &  0.35$\pm$0.02  & 72.96$\pm$0.22  & --  & --  & -- & 1751.96  & 1755.96 & 1766.84 &  0 & 0 \\
		Second-order  & -- & 72.22$^{+0.23}_{-0.23}$ & $-$0.16$^{+0.03}_{-0.03}$ & --  & -- & 1777.99 & 1781.99 & 1792.87 & 26.03 & 26.03 \\
		Third-order  & -- & 72.54$^{+0.27}_{-0.27}$ & $-$0.33$^{+0.06}_{-0.06}$ & 0.03$^{+0.29}_{-0.28}$ & -- & 1741.42 & 1747.42 & 1763.74 & $-$8.53 & $-$3.09 \\
		Forth-order  & -- & 72.90$^{+0.28}_{-0.29}$ & $-$0.50$^{+0.09}_{-0.07}$ &  1.66$^{+0.57}_{-0.83}$ & 2.86$^{+2.14}_{-2.29}$ & 1745.73 & 1753.73 & 1775.49 & $-$2.23 & 8.65 \\
		\hline\hline
	\end{tabular}
\end{table}

\begin{figure}[ht]
	\centering
	\includegraphics[width=0.45\textwidth]{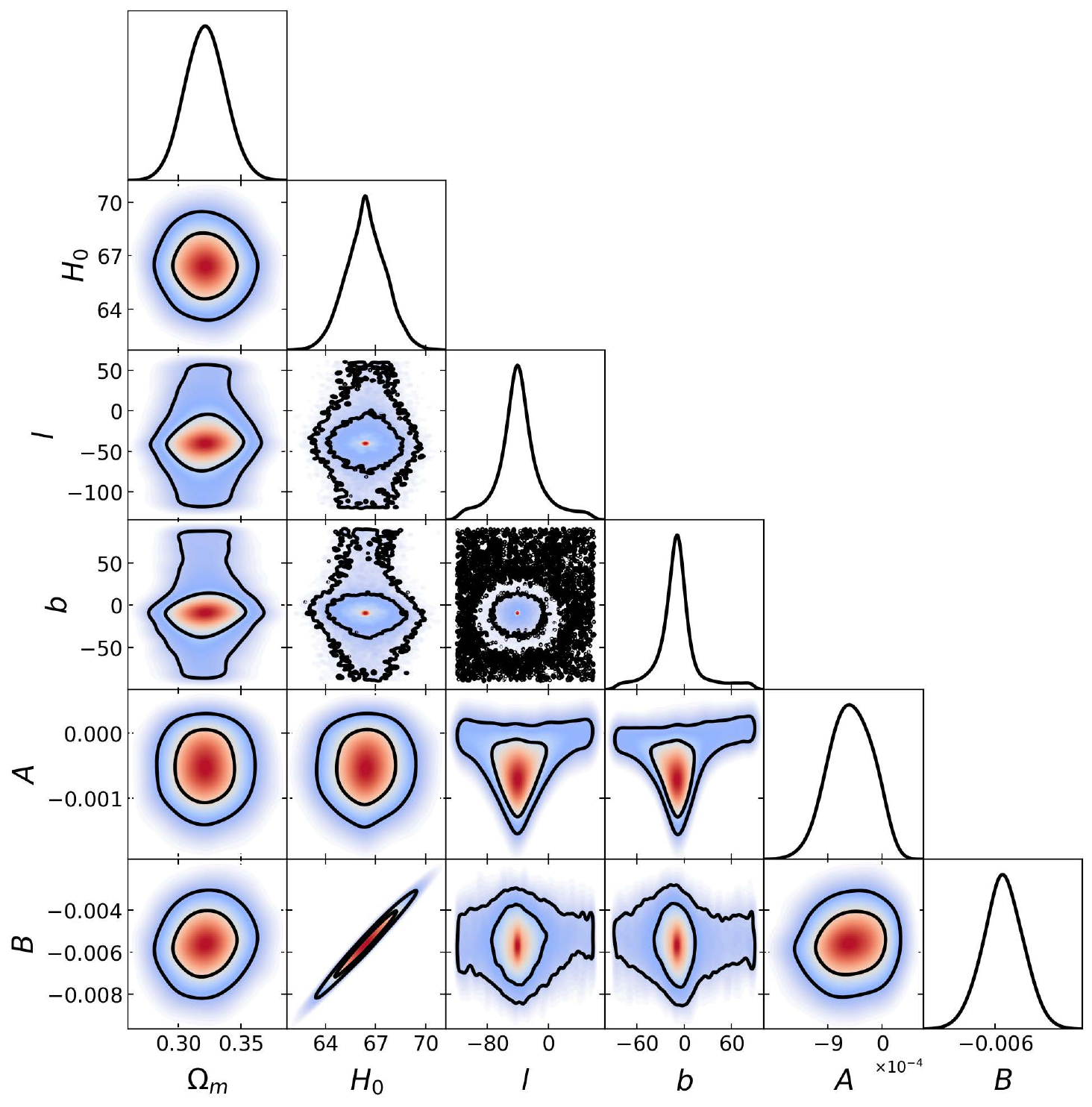}
	\includegraphics[width=0.45\textwidth]{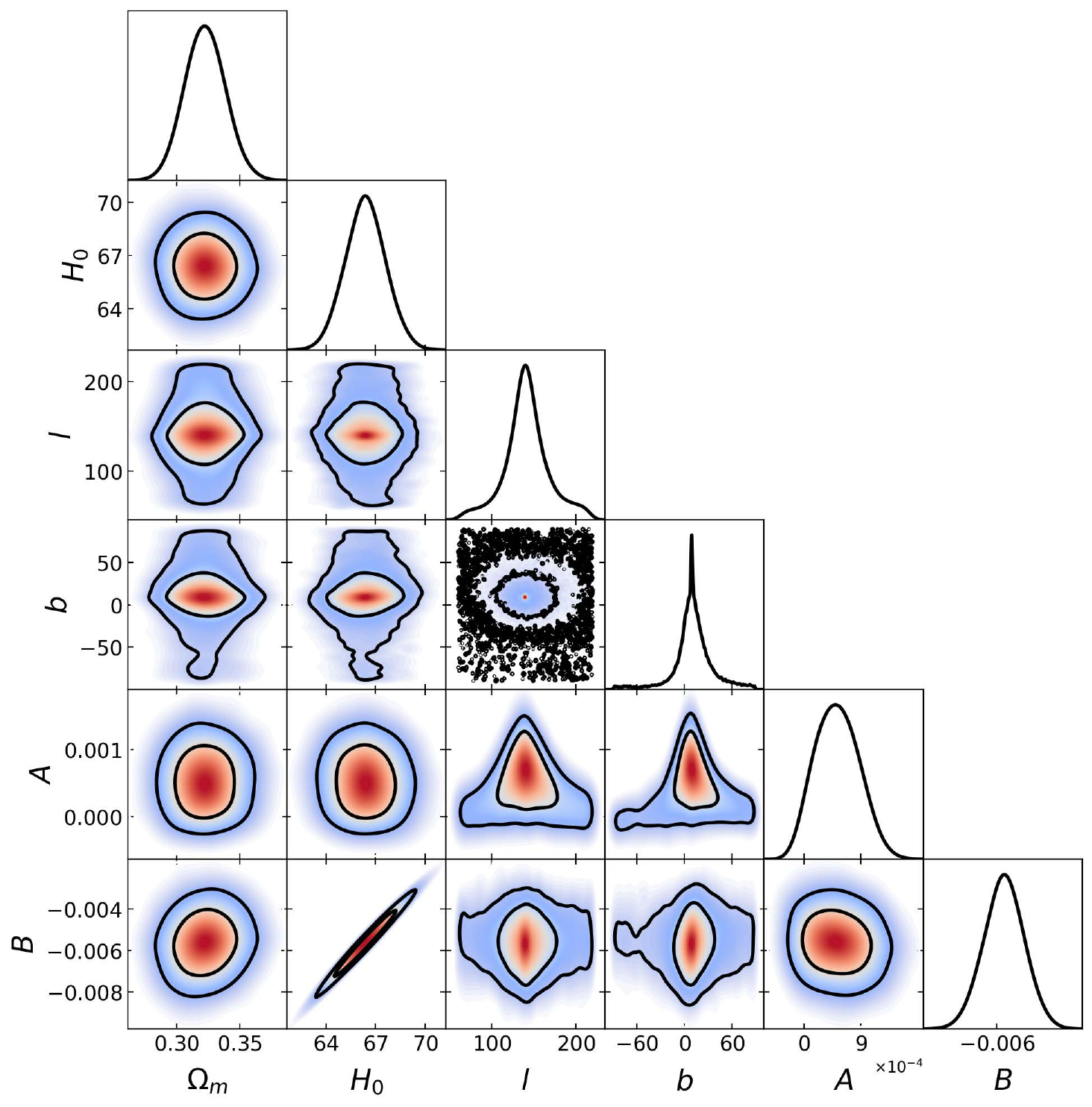}
	\caption{Confidence contours ($1\sigma$ and $2\sigma$) for the parameters space ($\Omega_{m}$, $H_{0}$, $l$, $b$, $A$ and $B$) from the joint dataset in the dipole model. The left panel shows $\Omega_{m}$ = 0.32$\pm$0.02, $H_{0}$ = 66.42$^{+1.23}_{-1.19}$ km/s/Mpc, $l$ = 319.79$^{\circ}$$^{+24.86}_{-21.20}$, $b$ = $-$9.30$^{\circ}$$^{+16.70}_{-18.51}$, $A$ = $-$5.2$\pm$3.6$\times$10$^{-4}$, $B$ = $-$5.6$\pm$1.0$\times$10$^{-3}$. The right panel shows $\Omega_{m}$ = 0.32$\pm$0.02, $H_{0}$ = 66.39$\pm$1.22 km/s/Mpc, $l$ = 140.50$^{\circ}$$^{+24.96}_{-21.14}$, $b$ = 10.06$^{\circ}$$^{+19.67}_{-15.09}$, $A$ = 5.2$^{+3.4}_{-3.8}$$\times$10$^{-4}$, $B$ = $-$5.6$\pm$1.1$\times$10$^{-3}$. }
	\label{snhz_dipole_2}  
\end{figure}

\begin{figure}[ht]
	\centering
	\includegraphics[width=0.45\textwidth]{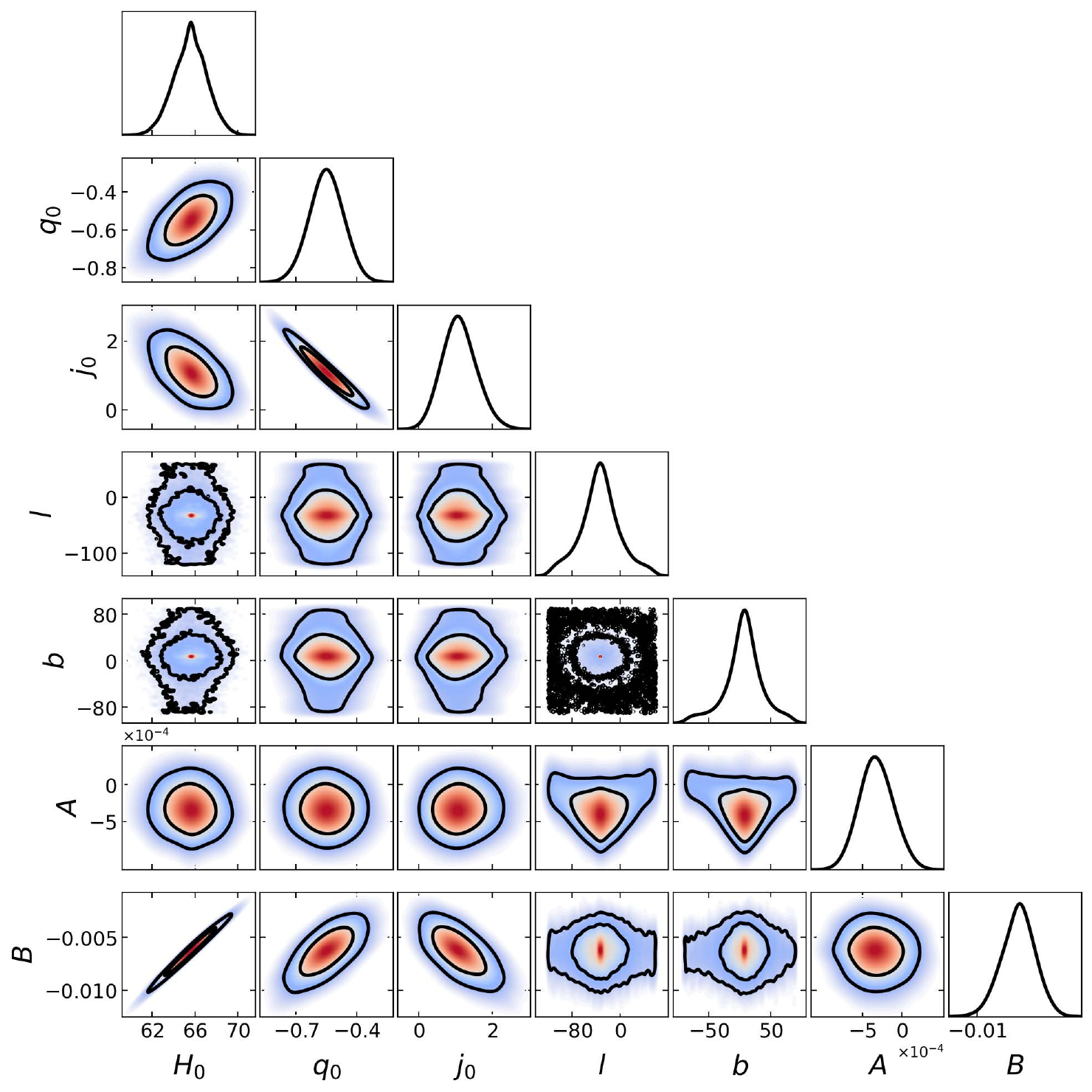}
	\includegraphics[width=0.45\textwidth]{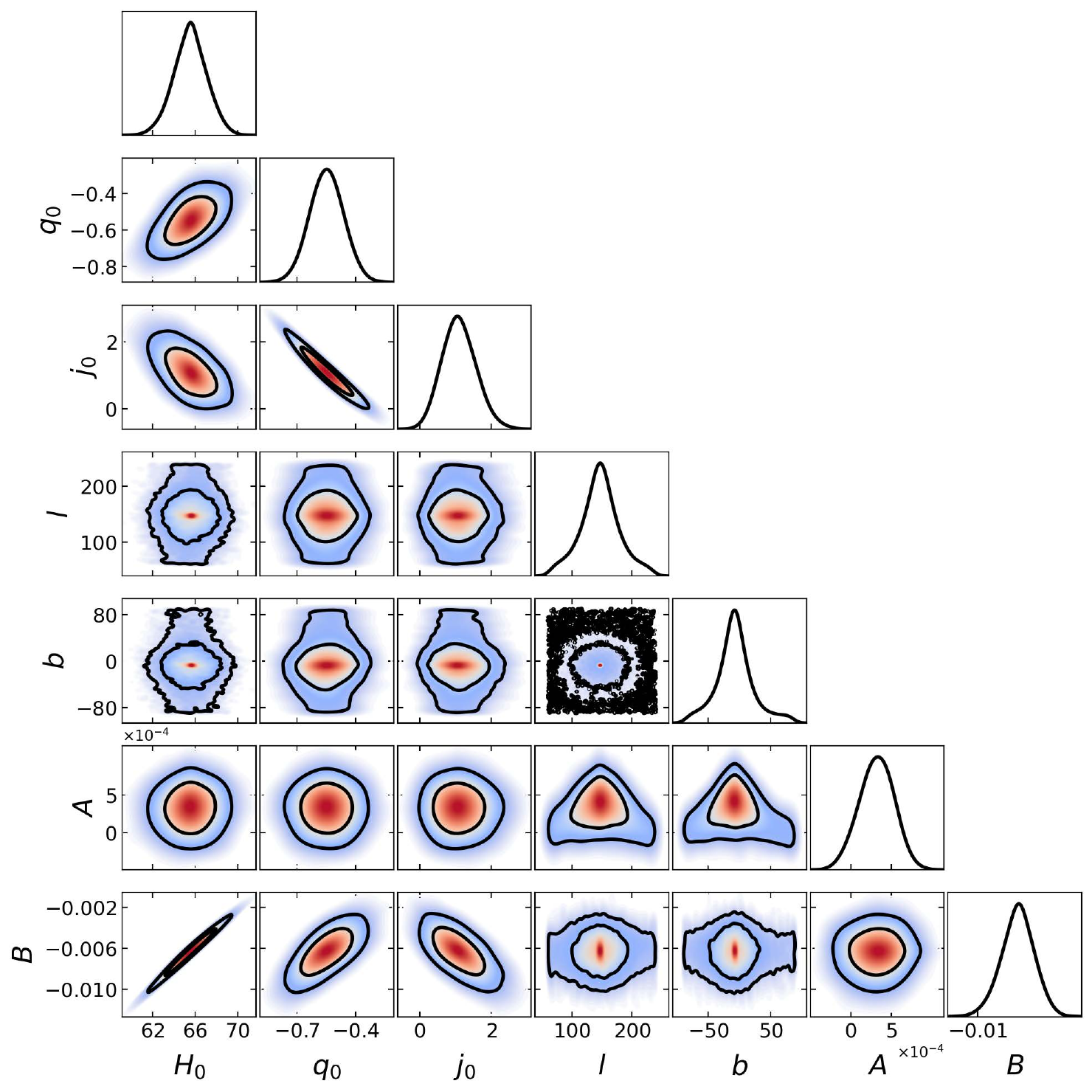}
	\caption{Confidence contours ($1\sigma$ and $2\sigma$) for the parameters space ($H_{0}$, $q_{0}$, $j_{0}$, $l$, $b$, $A$ and $B$) from the low redshift joint dataset using the third-order expansion. The constraints are ($l$ = 326.85$^{+30.24}_{-30.78}$, $b$ = 7.52$^{+24.38}_{-26.65}$, $A$ = $-$3.2$\pm$2.2$\times 10^{-4}$, $B$ = $-$6.3$\pm$1.5$\times 10^{-3}$) and ($l$ = 140.50$^{+24.96}_{-21.14}$, $b$ = 10.06$^{+19.67}_{-15.09}$, $A$ = 5.2$^{+3.4}_{-3.8}$$\times 10^{-4}$, $B$ = $-$5.6$\pm$1.1$\times 10^{-3}$), as shown in the left panel and the right panel respectively.}
	\label{z_dipole_2}       
\end{figure}

\label{lastpage}
\end{document}